\documentclass[aps,twocolumn]{revtex4}
\usepackage{epsfig}
\usepackage{graphicx}
\usepackage{amsmath,amssymb,amsfonts,bm}
\usepackage{array}
\usepackage{url}
\usepackage{hyperref}
\usepackage{multirow}
\usepackage{float}
\usepackage{lineno}
\usepackage{xspace}
\usepackage{ulem}
\usepackage[usenames,dvipsnames]{color}

\newcommand{\bef}{\begin{figure}}
\newcommand{\eef}{\end{figure}}
\newcommand{\bc}{\begin{center}}
\newcommand{\ec}{\end{center}}
\newcommand{\nn}{\nonumber}
\newcommand{\be}{\begin{equation}}
\newcommand{\ee}{\end{equation}}
\newcommand{\bea}{\begin{eqnarray}}
\newcommand{\eea}{\end{eqnarray}}

\def\ba{\begin{eqnarray}}
\def\ea{\end{eqnarray}}
\begin{document}
\title{Thermoelectric effects of an interacting hadron gas in the presence of an external magnetic field}
 
\author{Kamaljeet Singh}
\author{Kshitish Kumar Pradhan}
\author{Dushmanta Sahu}
\author{Raghunath Sahoo\footnote{Corresponding Author: Raghunath.Sahoo@cern.ch}}

\affiliation{Department of Physics, Indian Institute of Technology Indore, Simrol, Indore 453552, India}

\date{\today}

\begin{abstract}

The hot and dense hadronic medium formed during the heavy-ion collisions at the Relativistic Heavy Ion Collider and Large  Hadron Collider energies can show thermoelectric effects in the presence of temperature gradients and nonzero baryon chemical potential. In this article, we study the thermoelectric coefficients of an interacting hot and dense hadron gas using the relativistic Boltzmann transport equation under the relaxation time approximation. We discuss the thermoelectric properties within different frameworks of hardon resonance gas models. In the presence of an external magnetic field, the thermoelectric coefficients become anisotropic, which leads to Hall-like thermoelectric coefficients, namely Nernst coefficients, along with the magneto-Seebeck coefficients. For the first time, we also estimate the Thomson coefficient of the medium, which comes into the picture due to the temperature dependence of the Seebeck coefficient of the medium. In the context of studying the thermoelectric generator performance, we calculate the values of the thermoelectric figure of merit of the medium. 
\end{abstract}

\maketitle

\section{Introduction}

In heavy-ion collision experiments such as the Large Hadron Collider (LHC) or the Relativistic Heavy Ion Collider (RHIC), a deconfined state of nuclear matter can be produced at extremely high temperatures ($T$) and/or baryon chemical potentials ($\mu_{B}$). This thermalized state of matter is called quark-gluon plasma (QGP). QGP evolves~\cite{Brewer:2019oha, Busza:2018rrf} and cools down rapidly, producing a hot and dense hadronic matter that later freezes out leading to the detection of long-lived final-state particles. Studying the hadronic phase followed by the hadronization process is crucial for understanding the full evolution of heavy-ion collisions and extracting hidden information about the properties of dense nuclear matter. Due to the extreme energy density in the initial phase, the system continues to expand and cool down rapidly. Hence, the temperature gradients develop in both space and time. The temperature decreases with increased spatial distance from the collision point, which leads to radial temperature gradients. This temperature gradient plays a significant role in heavy-ion collisions because it, in turn, affects the evolution of the system. Currently, researchers are interested in understanding how the thermoelectric properties of the hot hadron gas influence the evolution of collision dynamics. 

The thermoelectric properties refer to the response of a material to a temperature gradient, which can generate an electric potential difference subjected to this temperature gradient~\cite{book}. A baryon current is produced in the medium at a finite baryon chemical potential. Similarly, a charge current will also be produced due to the presence of charges. Due to the baryon and charge currents in the medium, thermal and electrical transport properties are of significant interest in the study of heavy-ion collisions~\cite{Singh:2023pwf, Singh:2023ues}. The thermoelectric effects are proportional to the derivative of conductivity, and they are generally considered more sensitive to the responses to applied external fields. When thermoelectric coefficients are nonzero, temperature and chemical potential gradients in a conductor can drive an electric current. The Seebeck coefficient measures a material's efficiency in converting these gradients into electric current and is defined under “open circuit” conditions, where the net current is zero~\cite{Singh:2024emy}. This coefficient has been widely studied in condensed matter systems, including superconductors, quantum dot-coupled Majorana bound states, and various junctions and materials~\cite{PhysRevB.81.155457, PhysRevB.84.075420, PhysRevB.101.045101, PhysRevB.90.165115}. These thermoelectric transport properties of hot and dense hadron gas have been studied theoretically using various approaches such as relativistic kinetic theory~\cite{Bhatt:2018ncr, Das:2020beh, Das:2021qii}, hydrodynamics~\cite{Greif:2017byw, Prakash:1993bt, Kadam:2014xka}, and lattice QCD simulations~\cite{Meyer_2011, Kaczmarek_2022}. For some situations, the Seebeck coefficient is not constant but shows a gradient with the temperature gradient. When a current drives through this gradient, the Thomson coefficient can be observed, which describes how much heat is absorbed or released. The amount of heat exchange depends on the medium's properties and the direction of both the current and the temperature gradient.

In the peripheral heavy-ion collisions, a strong but transient magnetic field is produced due to the relativistic motion of the charged spectators. The order of this external magnetic field can reach up to $m_\pi^2 \sim 10^{18}$~G in Au-Au collisions at RHIC and \(15~m_\pi^2\) in Pb-Pb collisions at the LHC~\cite{Wang2021,MCINNES2016173}. Without any thermalized medium, this external magnetic field decays rapidly. But, in the presence of a thermalized medium, there is a finite electrical conductivity, which, by the Faraday law, produces an induced magnetic field in the direction of the external field~\cite{DEY2023122654}. This results in the external magnetic field decaying relatively slowly to sufficiently impact the thermodynamic and transport properties of the hot hadronic matter. The effect of magnetic fields on electrical and thermal conductivities has been explored thoroughly in literature~\cite{Singh:2023pwf,PhysRevD.106.034008,PhysRevD.104.094037}. This means that in the presence of a nonvanishing electromagnetic field, the thermoelectric effects will be modified, and one can estimate the magneto-Seebeck coefficient and the Nernst coefficient. Unlike the Seebeck coefficient, the Nernst coefficient is a Hall-type thermoelectric coefficient that becomes nonzero only in the presence of a finite magnetic field. When a conducting medium is exposed to a magnetic field, and a temperature gradient is applied perpendicular to it, the Nernst coefficient characterizes the resulting electric current, which is oriented normally to both the magnetic field and the temperature gradient~\cite{Singh:2024emy}. It has been previously explored in a hot and dense hadronic medium within the hadron resonance gas model~\cite{Bhatt:2021zui,Das:2020beh}. Moreover, in Ref.~\cite{Zhang:2024htn}, the authors have investigated the effects of magneto-thermoelectric properties on the diffusion of conserved charges.

In relativistic systems, the thermoelectric formalism differs notably from the nonrelativistic cases: here, thermal current requires a conserved number current, like the baryon number in QCD. This distinction affects the behavior of thermoelectric coefficients, as shown in studies of hadronic matter via the hadron resonance gas (HRG) model~\cite{Das:2021qii}. Studies in isotropic~\cite{Abhishek:2020wjm} and anisotropic~\cite{Kurian:2021zyb,Zhang:2020efz} QGP systems have also explored this, although using perturbative QCD, which is more applicable at very high temperatures. Near the QCD transition temperature (\(T_c\)), nonperturbative effects are relevant, as the QCD vacuum structure is nontrivial, influenced by chiral symmetry breaking and confinement-related condensates. However, understanding the thermoelectric properties of the hadronic medium is also of equal importance.
To understand the hot and dense hadronic matter formed in relativistic heavy-ion collisions, the ideal hadron resonance gas (IHRG) model is a widely used model. It explains the lattice QCD (lQCD) estimations of thermodynamic variables up to \(T \approx 150~\mathrm{MeV}\)~\cite{Bellwied:2013cta, Bellwied:2017ttj}, after which the hadrons start to melt, and the disagreements between the IHRG and lQCD appear. Unlike the lQCD, which breaks down at very high \(\mu_{B}\)~\cite{Borsanyi:2013bia, HotQCD:2014kol}, the IHRG model can be used to study highly dense hadronic or nuclear matter. However, the ideal HRG model is unable to explain the higher-order fluctuations of conserved charges, which have been predicted from the lQCD~\cite{Bazavov:2013dta, Bazavov:2017dus, Borsanyi:2018grb}. Thus, later on, better alternatives to the simplistic IHRG model have been explored extensively. One such widely used model is the excluded volume hadron resonance gas (EVHRG) model~\cite{Andronic:2012ut}. In this model, the hadrons have some finite radius and thus exert an outward pressure, which is the result of the excluded volume of the hadrons. The EVHRG model explains the thermodynamic properties and the conserved charged fluctuations of the hadronic matter very well, giving a good agreement with the lQCD data~\cite{Andronic:2012ut, Kadam:2015xsa, Kadam:2017iaz, Pal:2020ucy}. There are also other models, such as the repulsive mean-field hadron resonance gas (RMFHRG) model~\cite{Kapusta:1982qd, Olive:1980dy}, which uses a potential term to introduce the repulsive interaction between the hadrons. Recently, a new model has been explored called the van der Waals hadron resonance gas (VDWHRG) model~\cite{Vovchenko:2017cbu, Vovchenko:2016rkn}, which takes into account both attractive and repulsive interactions between the hadrons. It has been found that the VDWHRG model is the most reliable HRG model, which agrees with the lQCD estimations~\cite{Vovchenko:2016rkn, Samanta:2017yhh}. It also gives a second-order liquid-gas phase transition critical point at very high \(\mu_{B}\)~\cite{Vovchenko:2015vxa}, which can help study dense nuclear matter present in neutron stars.  

In this article, we offer a detailed examination of the thermoelectric properties and associated transport coefficients of the hot and dense hadronic medium in both the presence and the absence of an external magnetic field. We have used four different formalisms, namely, the IHRG, EVHRG, RMFHRG, and VDWHRG, to estimate the thermoelectric coefficients. Further, we also discuss the thermoelectric generator performance of the medium in terms of the thermoelectric figure of merit. This article is organized in the following manner. Section (\ref{sec: Formalism}) briefly gives the derivation of thermoelectric coefficients with and without an external magnetic field. Section (\ref{results}) discusses the results in detail and finally, in Sec. (\ref{summary}), we summarize the study with possible outlooks.
 
\label{intro}

\section{Formalism}\label{sec: Formalism}
In this section, we discuss the hadron resonance gas model and its extended versions that we use to calculate the thermoelectric coefficients of hot and dense hadron gas. We calculate these coefficients in both the absence and presence of an external magnetic field by using the Boltzmann transport equation under relaxation time approximation. For all our calculations, we have taken all the hadrons and their resonances up to the mass cutoff of 2.6 GeV from the Particle Data Group~\cite {ParticleDataGroup:2008zun}.

\subsection{Hadron resonance gas model}
The Hadron Resonance Gas (HRG) model is a theoretical framework used in high-energy nuclear physics to describe the thermodynamic properties of strongly interacting matter in the hadronic phase. The HRG model assumes that the hadronic matter is in thermal equilibrium, meaning that the temperature, chemical potential, and other thermodynamic variables are well-defined throughout the system. The various HRG models are discussed here.

\subsubsection{Ideal hadron resonance gas model}
The IHRG model assumes that a system of hadrons and resonances behaves as an ideal gas, meaning that the particles do not interact with each other and are considered to be point particles. 
The grand canonical partition function for an ideal hadron resonance gas can be written as~\cite{Pradhan:2022gbm},
\begin{equation}
\label{eq1}
ln Z^{id}_i = \pm \frac{Vg_i}{2\pi^2} \int_{0}^{\infty} k_i^2 dk_i\ ln\{1\pm \exp[-(\omega_i-\mu_i)/T]\}.
\end{equation}
Here, $g_i$ is the degeneracy, $k_i$ is the momentum, $m_i$ is the mass and $\omega_i = \sqrt{k_i^2 + m_i^2}$ is the energy of the $i{\rm th}$ hadron species. The $\pm$ sign corresponds to fermions and bosons, respectively.  $\mu_{i}$ denotes the corresponding chemical potential, which is given by 
\begin{equation}
\label{eq2}
\mu_i = b_i\mu_B ,
\end{equation}
where the baryon chemical potential is given by $\mu_{B}$ and $b_i$ denotes the baryon number of the $i{\rm th}$ hadron. Here, we have considered a simple case by taking the vanishing charge and strangeness chemical potential, i.e., $\mu_Q$ = $\mu_S$ = 0. The pressure $P_i$, energy density $\varepsilon_i$, and number density $n_i$ can now be obtained from the partition function, given as,
\begin{align}
\label{eq3}
P^{id}_i(T,\mu_i) &= \pm \frac{Tg_i}{2\pi^2} \int_{0}^{\infty} k_i^2 dk_i\ ln\{1\pm \exp[-(\omega_i-\mu_i)/T]\}, \\
\varepsilon^{id}_i(T,\mu_i) &= \frac{g_i}{2\pi^2} \int_{0}^{\infty} \frac{\omega_i\  k_i^2 dk_i}{\exp[(\omega_i-\mu_i)/T]\pm1},\\
n^{id}_i(T,\mu_i) &= \frac{g_i}{2\pi^2} \int_{0}^{\infty} \frac{k_i^2 dk_i}{\exp[(\omega_i-\mu_i)/T]\pm1}.
\end{align}

\subsubsection{Excluded volume hadron resonance gas model}
The EVHRG model is an extension of the IHRG model that takes into account the finite size of hadrons. Meanwhile, in an IHRG model, hadrons are treated as pointlike particles without considering their physical volume. The EVHRG model incorporates repulsive interactions by introducing an excluded volume for hadrons, preventing overlap, and thereby providing a more realistic description of hot and dense hadronic matter created in heavy-ion collisions.
In thermodynamically consistent excluded volume formulation, the transcendental equation for the pressure can be written as~\cite{Rischke:1991ke, Bhattacharyya:2013oya}

\be
P^{EV}(T,\mu)=P^{\text{id}}(T,\mu^{*}),
\label{prexcl}
\ee
where $\mu^{*}=\mu-vP^{EV}(T,\mu)$ is an effective chemical potential with $ v $ as the parameter 
corresponding to a proper volume of the particle.  At high temperatures and low densities, this prescription is equivalent to multiplying a
suppression factor of $\text{exp}(-vP^{EV}/T)$ to the pressure in the Boltzmann approximation. Therefore, the pressure  in the excluded volume hadron resonance gas model becomes

\be
P^{EV}(T,\mu)=e^{\frac{-vP^{EV}(T,\mu)}{T}}P^{{id}}(T,\mu),
\ee
where $P^{id}$ in Boltzmann approximation can be written as

\be
P^{{id}}(T,\mu)=\sum_{i}\frac{g_{i}}{2\pi^{2}}m_{i}^{2}T^{2}\mathcal{K}_{2}\bigg(\frac{m_{i}}{T}\bigg)\text{cosh}\bigg(\frac{\mu_{i}}{T}\bigg).
\ee
  $\mathcal{K}_{2}$ is the modified Bessel's function of second order. 
Other thermodynamical quantities can be readily obtained from Eq.~(\ref{prexcl}) by taking appropriate derivatives.  The number density and energy density, respectively, can be written as~\cite{Rischke:1991ke}
\be
n^{EV}(T,\mu)=\sum_{i}\frac{n^{id}_{i}(T,\mu^{*})}{1+\sum_{i}v_{i}n_{i}^{id}(T,\mu^{*})},
\ee
\be
\varepsilon^{EV}(T,\mu)=\sum_{i}\frac{\varepsilon^{id}_{i}(T,\mu^{*})}{1+\sum_{i}v_{i}n_{i}^{id}(T,\mu^{*})}.
\ee
\subsubsection{van der Waals hadron resonance gas model}
The van der Waals HRG model is a minimal-interaction extension of the IHRG model, which includes attractive interactions along with repulsive interactions due to the excluded volume of the hadrons. These interactions are assumed to exist between all pairs of baryons and all pairs of antibaryons. We incorporate repulsive interactions among the mesons, whereas the attractive interactions are assumed to be taken care of by considering the resonances in the model \cite{Pradhan:2022gbm, Dashen:1969ep}. The baryon-antibaryon and meson-(anti)baryon VDW interactions are neglected. The van der Waals equation of state can be written as \cite{Pradhan:2022gbm, Samanta:2017yhh}
 \begin{equation}
\label{eq7}
    \Bigg( P + \bigg(\frac{N}{V}\bigg)^{2}\rm a\Bigg)\big(V-N \rm b\big) =  N T,
\end{equation}
where parameters $a$ and $b$ (positive) are the VDW parameters that describe attractive and repulsive interactions. The system's pressure, volume, temperature, and number of particles are denoted by $P$, $V$, $T$, and $N$, respectively.

One can simplify the above equation in terms 
 of the number density, $n \equiv N/V$ as
\begin{equation}
\label{eq8}
    P(T,n) = \frac{nT}{1-{\rm{b}} n}- {\rm{a}} n^{2},
\end{equation}
The repulsive interactions are included in the first term of Eq.~(\ref{eq8}) by replacing the total volume $V$ with the effective volume available to particles using the proper volume parameter $b = 16\pi r^{3}/3$, $r$ being the particle hardcore radius. The second term takes care of the attractive interactions between particles.

The VDW equation of state in the grand canonical ensemble can then be written as \cite{Samanta:2017yhh,Vovchenko:2015pya} 
\begin{equation}
\label{eq9}
    P(T,\mu) = P^{id}(T,\mu^{*}) - {\rm{a}}n^{2}(T,\mu),
\end{equation}
where the $n(T,\mu)$ is the particle number density of the VDW hadron gas and is given by
\begin{equation}
\label{eq10}
    n(T,\mu) = \frac{\sum_{i}n_{i}^{id}(T,\mu_{i}^{*})}{1+{\rm b}\sum_{i}n_{i}^{id}(T,\mu_{i}^{*})}.
\end{equation}
Here, $\mu^{*}$ is the modified chemical potential given by 
\begin{equation}
\label{eq11}
    \mu^{*} = \mu - {\rm b}P(T,\mu) - {\rm{ab}}n^{2}(T,\mu) + 2{\rm a}n(T,\mu).
\end{equation}
Now, energy density $\varepsilon(T,\mu)$ can be obtained as,
\begin{equation}
\label{eq13}
\varepsilon(T,\mu) = \frac{\sum_{i}\varepsilon_{i}^{id}(T,\mu_{i}^{*})}{1+{\rm b}\sum_{i}n_{i}^{id}(T,\mu_{i}^{*})} - {\rm a}n^{2}(T,\mu).
\end{equation}

The total pressure in the VDWHRG model comes from the mesons, baryons, and antibaryons separately because of the interaction and is given by \cite{Samanta:2017yhh, Vovchenko:2016rkn} 
\begin{equation}
\label{eq14}
P(T,\mu) = P_{M}(T,\mu) + P_{B}(T,\mu) + P_{\bar{B}}(T,\mu),
\end{equation}
where the $P_{M}(T,\mu), P_{B(\bar B)}(T,\mu)$ are the contributions to pressure from mesons and (anti)baryons, respectively and are given by,
\begin{align}
\label{eq15}
P_{M}(T,\mu) &= \sum_{i\in M}P_{i}^{id}(T,\mu^{*M}),  \\    
P_{B}(T,\mu) &= \sum_{i\in B}P_{i}^{id}(T,\mu^{*B})-{\rm a}n^{2}_{B}(T,\mu),\\
P_{\bar{B}}(T,\mu) &= \sum_{i\in \bar{B}}P_{i}^{id}(T,\mu^{*\bar{B}})-{\rm a}n^{2}_{\bar{B}}(T,\mu).
\end{align}
Here, $M$, $B$, and $\bar B$ represent mesons, baryons, and antibaryons. $\mu^{*M}$ is the modified chemical potential of mesons because of the excluded volume correction ($b_M$), and $\mu^{*B}$ and $\mu^{*\bar B}$ are the modified chemical potentials of baryons and antibaryons due to VDW interactions ($a$ and $b$)~\cite{Sarkar:2018mbk}. The modified chemical potential for mesons and (anti)baryons can be obtained from Eq.~(\ref{eq2}) and Eq.~(\ref{eq11}) as, 
\begin{align}
\label{eq18}
\mu^{*M} &= -{\rm b}_MP_{M}(T,\mu),\\
\mu^{*B(\bar B)} &= \mu^{B(\bar B)}-{\rm b}_B P_{B(\bar B)}(T,\mu)-{\rm{ab}}_B n^{2}_{B(\bar B)}+2{\rm a}n_{B(\bar B)},
\end{align}
where $n_{M}$, $n_{B}$ and $n_{\bar B}$ are the modified number densities of mesons, baryons and antibaryons, respectively, which are given by
\begin{align}
\label{eq20}
    n_{M}(T,\mu) &= \frac{\sum_{i\in M}n_{i}^{id}(T,\mu_{i}^{*M})}{1+{\rm b}_M\sum_{i\in M}n_{i}^{id}(T,\mu_{i}^{*M})},\\
    n_{B(\bar B)}(T,\mu) &= \frac{\sum_{i\in B(\bar B)}n_{i}^{id}(T,\mu_{i}^{*B(\bar B)})}{1+{\rm b}_B\sum_{i\in B(\bar B)}n_{i}^{id}(T,\mu_{i}^{*B(\bar B)})}.
\end{align}
\subsubsection{Repulsive mean-field hadron resonance gas model}
 The RMF corrections are often implemented within a mean-field approximation, where the repulsive interactions are treated as an effective mean field acting on each hadron. This mean field modifies the single-particle energies of the hadrons, affecting their thermodynamic properties such as the equation of state, pressure, and density.
 A repulsive mean-field approach is utilized to incorporate the effects of repulsive interactions among hadrons, 
as it was used in Refs. \cite{ Kapusta:1982qd,Olive:1980dy} 
and recently, in the case of baryons in Ref.\cite{Huovinen:2017ogf}.
In this approach, it is assumed that the repulsive interactions lead to a shift in the single-particle energy and is given by
\be
\tilde \omega_{i}=\sqrt{\vec{k_{i}}^2+m_{i}^2}+U(n)=\omega_{i}+U(n),
\label{dispersion}
\ee
 where the potential energy $U$ characterizes the repulsive interaction between hadrons and depends on the total hadron density $n$. For a given hadron potential $V({\bf{r}})$, the potential energy can be expressed as $U(n)=Kn$. The phenomenological parameter $K$ is determined by integrating the potential $V(\bf r)$ over the spatial volume \cite{Kapusta:1982qd,Olive:1980dy}.

In this work, we assign different repulsive interaction parameters for baryons and mesons. We denote the mean-field parameter 
for baryons ($B$) and antibaryons ($\bar{B}$) by $K_B$, while for mesons we denote it by  $K_M$. Thus,  for baryons (antibaryons)
\be
U(n_{B\{\bar{B}\}})=K_Bn_{B\{\bar{B}\}},
\label{potenbar}
\ee
and for mesons
\be
U(n_M)=K_Mn_M.
\label{potenmes}
\ee
The repulsion parameter $K_B$ = 0.450 GeV fm$^3$ is taken the same for all (anti)baryons and $K_M$ = 0.050 GeV fm$^3$ for all mesons as in Ref.~\cite{Huovinen:2017ogf}.
The total hadron number density is
\be
n(T,\mu)=\sum_{a}n_{a}=n_B+n_{\bar{B}}+n_M,
\ee
where $n_{i}$ is the number density of $i{\rm th}$ hadronic species. Note that $n_B$, $n_{\bar{B}}$, and $n_M$ are total baryon,
 antibaryon and meson number densities, respectively. Explicitly, for baryons,
\be
n_{B}=\sum_{i\in B}\frac{g_{i}}{2\pi^2}\int_{0}^{\infty}\:\frac{k^{2}_{i}dk_{i}}{e^{\big(\frac{\omega_{i}-\mu_{\text{eff}}^i}{T}\big)}+1},
\label{numdenbaryon}
\ee
where the sum is over all the baryons. Here, $\mu_{\text{eff}}^i=b_i\mu_{B}-U(n_{B})$
is the baryon effective chemical potential, with $b_i$ being the baryonic number of $i{\rm th}$ baryon and $\mu_B$, 
the baryon chemical potential. Similarly, for antibaryons
\be
n_{\bar{B}}=\sum_{i\in \bar{B}}\frac{g_{i}}{2\pi^2}\int_{0}^{\infty}\:\frac{k^{2}_{i}dk_{i}}{e^{\big(\frac{\omega_{i}-\bar{\mu}_{\text{eff}}^i}{T}\big)}+1},
\label{numdenantbaryon}
\ee
 where $\bar{\mu}_{\text{eff}}^{i}=(\bar b_i\mu_B-U(n_{\bar{B}}))$ is an antibaryon effective chemical potential with $\bar b_i=-b_i$. For mesons,
\be
n_{M}=\sum_{i\in M}\frac{g_{i}}{2\pi^2}\int_{0}^{\infty}\:\frac{k^{2}_{i}dk_{i}}{e^{\big(\frac{\omega_i-K_Mn_M}{T}\big)}-1},
\label{numdenmeson}
\ee
where the sum is over all the mesons. Note that $\mu=0$ for mesons since the baryon charge is zero for them. 
The total (anti)baryon energy density  is
\begin{align}
\varepsilon_{B\{\bar{B}\}}=&\sum_{i\in B\{\bar{B}\}}\frac{g_{i}}{2\pi^2}\int_{0}^{\infty}\:\frac{k^{2}_{i}dk_{i}\tilde \omega_{i}}{e^{\big(\frac{\omega_{i}-\mu^i_{\text{eff}}\{\bar{\mu}^i_{\text{eff}}\}}{T}\big)}+1}~ \nn\\
&+\phi_{B\{\bar{B}\}}(n_{B\{\bar{B}\}}),\nonumber\\
= &\sum_{i\in B\{\bar{B}\}}\frac{g_{i}}{2\pi^2}\int_{0}^{\infty}\:\frac{k^{2}_{i}dk_{i}\tilde \omega_{i}}{e^{\big(\frac{\tilde\omega_{i}-b_i\mu^i_{B}\{\bar{\mu}^i_{B}\}}{T}\big)}+1}~ \nn\\
&+\phi_{B\{\bar{B}\}}(n_{B\{\bar{B}\}}),
\label{endenbaryon}
\end{align}
and for mesons
\be
\varepsilon_{M} = \sum_{i\in{M}}\frac{g_{i}}{2\pi^2}\int_{0}^{\infty}\:\frac{k^{2}_{i}dk_{i}\tilde \omega_{i}}{e^{\frac{\tilde \omega_{i}}{T}}-1}+\phi_{M}(n_M),
\label{endenmeson}
\ee
where $\phi(n)$ represents the energy density correction to avoid double-counting the potential. 
One can determine it by using the condition that $\tilde \omega_{i}=\frac{\partial \varepsilon}{\partial n_{i}}$. After doing the 
derivative of baryon energy density with respect to baryon net number density and using Eq. (\ref{potenbar}), we get

\be
\frac{\partial \phi_{B\{\bar{B}\}}}{\partial n_{B\{\bar{B}\}}}=-K_Bn_{B\{\bar{B}\}},
\ee

and hence

\be
\phi_B(n_{B\{\bar{B}\}})=-\frac{1}{2}K_Bn_{B\{\bar{B}\}}^2.
\ee

Also, for mesons, one can obtain in a similar way
\be
\phi_M(n_M)=-\frac{1}{2}K_Mn_M^2.
\ee

The pressure of the gas can now be readily obtained. For baryons
\begin{align}
P_{B\{\bar{B}\}}(T,\mu)=T\sum_{i\in B\{\bar{B}\}}\frac{g_{i}}{2\pi^2}\int_{0}^{\infty}k_{i}^{2}dk_{i}\text{ln}\bigg[1+\nn\\ 
e^{-\big(\frac{\omega_i-\mu^i_{\text{eff}}\{\bar{\mu}^i_{\text{eff}}\}}{T}\big)}\bigg]
-\phi_{B\{\bar{B}\}}(n_{B\{\bar{B}\}}),
\end{align}
and for mesons
\be
P_M(T)=T\sum_{i\in M}\frac{g_{i}}{2\pi^2}\int_{0}^{\infty}k^{2}_{i}dk_{i}\text{ln}\bigg[1+ e^{-(\frac{\tilde \omega_i}{T})}\bigg]-\phi_M(n_M).
\ee
\subsection{Thermoelectric coefficients in the absence of magnetic field.}
\label{formalism1}
 To study the thermoelectric coefficients of hot and dense hadron gas in the absence of a magnetic field, we consider the Boltzmann transport equation (BTE) under relaxation time approximation (RTA). Under RTA, the Boltzmann equation can be interpreted as a linear expansion of the single-particle total distribution function ($f_i$) around the single-particle equilibrium distribution function ($f_i^{0}$). Here, ($f_i$) can be written as $f_i=f^0_i+\delta f_i$, where ($\delta f_i$) represents the deviation from equilibrium. The total single-particle distribution function for $i{\rm th}$ species at equilibrium, is given by
\begin{align}\label{Dis-f}
f^0_i= \frac{1}{e^{\frac{\omega_i-{\rm b}_i\mu_B}{T}}\pm 1}~,
\end{align}
where $\omega_i=\sqrt{\vec{k_i}^2+m_i^2}$ is the single particle energy,  $\mu_B$ is the baryon chemical potential, $b_i$ denotes the baryon number of $i{\rm th}$ particle, e.g. for baryons $b_i=1$, for antibaryons $b_i=-1$ and for mesons $b_i=0$. The $\pm$ sign stands for fermions and bosons, respectively. In local thermodynamic equilibrium, the spatial dependence of the distribution functions appears due to spatial gradients of temperature and baryon chemical potential. The linearized BTE under RTA in the local rest frame (LRF), for particle species $i$ can be written as \cite{Das:2020beh, Singh:2023pwf},

\begin{equation}
\frac{\partial f_i}{\partial t} + \vec{v}_i.\vec{\nabla}f_i+q_i\vec{E}.\frac{\partial f_i}{\partial\vec{k_i}} = 
  -\frac{\delta f_i(\vec{x_i},\vec{k_i})}{\tau^{i}_R(\vec{k_i})} ,
 \label{equnew1}
\end{equation}
where $\tau^{i}_R$ denotes the relaxation time of the particle species $i$. The equilibrium distribution function satisfies, 
\begin{align}
 \frac{\partial f^{0}_{i}}{\partial \vec{k_i}}=\vec{v}_i\frac{\partial f^{0}_{i}}{\partial \omega_i},
 ~~ \frac{\partial f^{0}_{i}}{\partial \omega_i}=-\frac{f^{0}_{i}(1\mp f^{0}_{i})}{T}, 
\label{equnew2}
 \end{align}
  $\vec{v}_i=\vec{k_i}/\omega_i$ is the velocity of the particle. The gradient of the equilibrium distribution function $\vec{\nabla}f^{0}_{i}$ can be expressed as,
\begin{align}
 \vec{\nabla}f^{0}_{i} = T \bigg[\omega_i\vec{\nabla}\left(\frac{1}{T}\right)-b_i\vec{\nabla}\left(\frac{\mu_B}{T}\right)\bigg]\frac{\partial f^{0}_{i}}{\partial\omega_i}.
 \label{equnew3}
\end{align}

Using the Gibbs-Duhem relation, we then have,
\begin{align}
 \vec{\nabla}f^{0}_{i} = - \frac{\partial f^{0}_{i}}{\partial\omega_i}\bigg(\omega_i-b_i h\bigg)\frac{\vec{\nabla}T}{T}.
\label{equnew4}
 \end{align}
 where $h=\frac{\varepsilon + P}{n}$ is the enthalpy per particle, $\varepsilon$, $P$, and $n$ are total energy density, total pressure, and net baryon density of the system, respectively.
 With leading-order contribution, we can consider an ansatz of $\delta f_i$ as~\cite{Singh:2023pwf, Gavin:1985ph}
\begin{align}\label{delta-f0}
	\delta f_i = (\vec{k_i} \cdot \vec{\Omega}) \frac{\partial f^0_i}{\partial \omega_i}~.
\end{align}  
The general form of unknown vector ${\vec \Omega}$ can be assumed as a linear combination of perturbing forces leading the medium out of thermal equilibrium as \begin{align}\label{Omega00}
\vec{\Omega} = &~\alpha_1 \vec{E} +\alpha_2 \vec{\nabla}T ~.
\end{align}  
The unknown coefficients $\alpha_j$ ($j=1,2$) determine the strength of the respective gradient force fields driving the system away from equilibrium.
Using Eq.\eqref{equnew4} and Eq.\eqref{equnew2} in Eq.\eqref{equnew1}, we can write the deviation of the equilibrium distribution function as,
\begin{align}
 \delta f_i = -\tau^{i}_R\frac{\partial f_i^{0}}{\partial \omega_i}\bigg[q_i(\vec{E}.\vec{v}_i)-\bigg(\frac{\omega_i-b_i h}{T}\bigg)\vec{v}_i.\vec{\nabla} T\bigg].
 \label{equnew5}
\end{align}
According to kinetic theory, the electric current $(\vec{j})$ of the system can be defined
using the deviation from the equilibrium distribution function $\delta f_i$ as,
\begin{align}
 \vec{j}& = \sum_i g_i\int \frac{d^3k_i}{(2\pi)^3}q_i\vec{v}_i\delta f_i\nonumber\\
 & = \sum_i \frac{g_i}{3} \int \frac{d^3k_i}{(2\pi)^3}\tau^{i}_R q_i^2 v^2_i\bigg(-\frac{\partial f_i^{0}}{\partial \omega_i}\bigg)\vec{E}\nonumber\\
 & -\sum_i \frac{g_i}{3} \int \frac{d^3k_i}{(2\pi)^3}\tau^{i}_Rq_iv^2_i
 \bigg(\frac{\omega_i-b_i h}{T}\bigg)\bigg(-\frac{\partial f_i^{0}}{\partial \omega_i}\bigg)\vec{\nabla} T.
 \label{equnew6}
\end{align}
In the above equation, we have used $\langle v^l_iv^j_i\rangle=\frac{1}{3}v_i^2\delta^{lj}$. Here, the sum is over all the baryons, antibaryons, and mesons. 
Also, for a relativistic system, the thermal current is defined with reference to the conserved baryon current. The thermal current arises when energy flows relative to the baryonic enthalpy. 
Hence, the heat current of the hot hadronic medium can be defined as \cite{Gavin:1985ph},

\begin{align}
\label{heat_current}
 \vec{I} & = \sum_i g_i\int \frac{d^3k_i}{(2\pi)^3}k_if_i - h \sum_i b_i g_i\int\frac{d^3k_i}{(2\pi)^3} v_i f_i\nonumber\\
 & = \sum_i g_i\int \frac{d^3k_i}{(2\pi)^3}\frac{k_i}{\omega_i}\left(\omega_i-b_i h\right)\delta f_i.
 \end{align}
 Use Eq. \ref{equnew5} with \ref{heat_current}
 \begin{align}
 \vec{I}& = \sum_i \frac{g_i}{3} \int \frac{d^3k_i}{(2\pi)^3} \tau^{i}_R q_i v_i^2\left(\omega_i-b_i h\right)\left(-\frac{\partial f_i^{0}}{\partial\omega_i}\right)\vec{E}\nonumber\\
 & -\sum_i \frac{g_i}{3T} \int \frac{d^3k_i}{(2\pi)^3} \tau^{i}_R v_i^2\left(\omega_i-b_i h\right)^2\left(-\frac{\partial f_i^{0}}{\partial\omega_i}\right)\vec{\nabla}T.
 \label{equnew9}
\end{align}
We can define the Seebeck coefficient $S$ using Eq.\eqref{equnew6} by setting $\vec{j}=0$ so that the electric field and temperature gradient become proportional to each other. Here, the proportionality factor is known as the Seebeck coefficient \cite{Singh:2024emy}. Hence from  Eq.\eqref{equnew6} we get, 
\begin{align}
 \vec{E}=S\vec{\nabla}T,
\end{align}
and hence, 
\begin{align}
 S &= \frac{\sum_i \frac{g_i}{3}\int \frac{d^3k_i}{(2\pi)^3}\tau^{i}_R q_i v_i^2\left(\omega_i-b_ih\right)\left(-\frac{\partial f_i^{0}}{\partial\omega_i}\right)}{T\sum_i \frac{g_i}{3}\int \frac{d^3k_i}{(2\pi)^3}\tau^{i}_R q^2_i v_i^2\left(-\frac{\partial f_i^{0}}{\partial\omega_i}\right)}\nonumber\\
 &=\frac{\sum_i \frac{g_i}{3T}\int \frac{d^3k_i}{(2\pi)^3}\tau^{i}_R q_i \left(\frac{\vec{k_i}}{\omega_i}\right)^2\left(\omega_i-b_ih\right)f^{0}_{i}(1\mp f^{0}_{i})}{T\sum_i \frac{g_i}{3T}\int \frac{d^3k_i}{(2\pi)^3}\tau^{i}_R q^2_i \left(\frac{\vec{k_i}}{\omega_i}\right)^2f^{0}_{i}(1\mp f^{0}_{i})}\nonumber\\
 &= \frac{\mathcal{I}_{1}/T^2}{\sigma_{el}/T}.
 \label{equnew11}
\end{align}
where, the electrical conductivity $\sigma_{el}$ can be identified from Eq.\eqref{equnew6} as,
\begin{align}
 \sigma_{el}  & = \sum_i \frac{g_i}{3T}\int \frac{d^3k_i}{(2\pi)^3}\tau^{i}_R q^2_i \left(\frac{\vec{k_i}}{\omega_i}\right)^2f^{0}_{i}(1\mp f^{0}_{i}),
\end{align}
and the integral $\mathcal{I}_{1}$ in Eq.\eqref{equnew11} is, 
\begin{align}
 \mathcal{I}_{1} = \sum_i \frac{g_i}{3T}\int \frac{d^3k_i}{(2\pi)^3}\tau^{i}_R q_i \left(\frac{\vec{k_i}}{\omega_i}\right)^2\left(\omega_i-b_ih\right)f^{0}_{i}(1\mp f^{0}_{i}). 
 \label{I31equ}
\end{align}
Here, the Seebeck coefficient can be both positive or negative because the numerator depends linearly on an electric charge while the integrand itself is not necessarily positive definite. As the Seebeck coefficient ($S$) is temperature dependent, the Thomson effect originates in the medium. The Thomson coefficient ($Th$) is related to the Seebeck coefficient as
\begin{align}\label{Thomson}
    Th = T\frac{dS}{dT}.
\end{align}
The above relation is usually known as the first Thomson relation, and it can be derived from energy conservation~\cite{PhysRevLett.125.106601}. The electric current and heat current can modify due to these thermoelectric coefficients as,
\begin{align}
 \vec{j}&=\sigma_{el}\vec{E}-\sigma_{el}S \vec{\nabla}T. \label{equnew14a}\\
 \vec{{I}}&=T\sigma_{el}S\vec{E}-\kappa_0\vec{\nabla}T,
 \label{equnew14}
\end{align}
where $\kappa_0$ is the coefficient of the thermal conductivity and is expressed as \cite{Singh:2023pwf},
\begin{align}
 \kappa_0=\sum_i\frac{g_i}{3T^2}\int\frac{d^3k_i}{(2\pi)^3}\tau^{i}_R\left(\frac{\vec{k_i}}{\omega_i}\right)^2\left(\omega_i-b_ih\right)^2f^{0}_{i}(1\mp f^{0}_{i}).
 \label{equnew15}
\end{align}
Using Eq.\eqref{equnew14a} and Eq.\eqref{equnew14}, we can express the heat current $\vec{{I}}$ in terms  of electric current $\vec{j}$ in the following way,
\begin{equation}
 \vec{{I}}=TS\vec{j}-\left(\kappa_0-T\sigma_{el}S^2\right)\vec{\nabla}T.
 \label{equnew16}
\end{equation}
The three transport coefficients, namely $S$, $\sigma_{el}$, and $\kappa_0$, are closely related to each other because of the common factors such as mobility and concentration of medium constituents. The thermoelectric performance of any thermoelectric material can be measured by using the dimensionless quantity named as the figure of merit ($ZT$), given as~\cite{Nemir2010}
\begin{align}
\label{eqnZT}
    ZT = \frac{S^{2}\sigma_{el}T}{\kappa_{0}}.
\end{align}
 It is to be noted that mesons contribute through the total enthalpy of the system as well as in the total electrical conductivity of the system, which enters the denominator of Eq. \ eqref {equnew11}. The presence of a magnetic field makes this picture more complicated, which we discuss in the next section.

\subsection{Thermoelectric coefficients in the presence of magnetic field.}
\label{formalism2}
The RBTE for a single hadron species under RTA in the presence of an electromagnetic field can be expressed as, 
\begin{align}
 \frac{\partial f_i}{\partial t} +\vec{v}_i.\frac{\partial f_i}{\partial \vec{x_i}}+q_i\left(\vec{E}+\vec{v}_i\times\vec{B}\right).\frac{\partial f_i}{\partial\vec{k_i}} = -\frac{\delta f_i}{\tau^{i}_R},
 \label{equnew21}
\end{align}

To solve the RBTE as given in Eq.\eqref{equnew21} we take an ansatz to express the deviation of the distribution function from the equilibrium in the following way~\cite{Gavin:1985ph}
  \begin{align}
  \delta f_i = (\vec{k_i}.~\vec{\Omega})\frac{\partial f^{0}_{i}}{\partial\omega_i},
\label{equnew22}
  \end{align}
with $\vec{\Omega}$ being related to a temperature gradient, electric field, the magnetic field and in general, can be written as
\begin{align}
 \vec{\Omega}= &\alpha_1 \vec{E}+\alpha_2\vec{B}+\alpha_3(\vec{E}\times\vec{B})+\alpha_4\vec{\nabla}T+ \alpha_5(\vec{\nabla}T\times\vec{B})\nonumber\\
 &~~~~~~~~~~~~~~~~~~~~~~~~~~~~~~~~~+\alpha_6(\vec{\nabla}T\times \vec{E}).
 \label{equnew23}
\end{align}
 Using Eqs.\eqref{equnew22} and \eqref{equnew23} RBTE as given in  Eq.\eqref{equnew21} can be expressed as,
\begin{align}
&\vec{v}_i.\bigg[- \frac{\partial f^{0}_{i}}{\partial\omega_i}\bigg(\frac{\omega_i-b_ih}{T}\bigg)\vec{\nabla}T\bigg]+q_i(\vec{E}.\vec{v}_i)\frac{\partial f^{0}_{i}}{\partial \omega_i}\nonumber\\
&-q_i \vec{v}_i.(\vec{\Omega}\times\vec{B})\frac{\partial f^{0}_{i}}{\partial \omega_i} = -\frac{\omega_i}{\tau^{i}_R}(\vec{v}_i.\vec{\Omega})\frac{\partial f^{0}_{i}}{\partial \omega_i}.
\label{equnew24}
\end{align}
By using $\vec{\Omega}$ as given in Eq.\eqref{equnew23}, RBTE as given in Eq.\eqref{equnew24}, can be expressed as~\cite{Singh:2024emy},
\begin{align}
 &q_i(\vec{E}.\vec{v}_i)-\alpha_1 q_i\vec{v}_i.(\vec{E}\times\vec{B})-\alpha_3 q_i  (\vec{E}.\vec{B})(\vec{v}_i.\vec{B})+\alpha_3 \nonumber\\
 &q_i (\vec{v}_i.\vec{E})-\alpha_4 q_i  \vec{v}_i.(\vec{\nabla}T\times \vec{B})-\alpha_5q_i(\vec{\nabla}T.\vec{B})(\vec{v}_i.\vec{B})\nonumber\\
 &+\alpha_5q_i(\vec{v}_i.\vec{\nabla}T)-\alpha_6q_i(\vec{\nabla}T.\vec{B})(\vec{v}_i.\vec{E})+\alpha_6q_i(\vec{E}.\vec{B})(\vec{v}_i.\vec{\nabla}T)\nonumber\\
 &-\bigg(\frac{\omega_i-b_ih}{T}\bigg)(\vec{v}_i.\vec{\nabla} T)
  = -\frac{\omega_i}{\tau^{i}_R}\bigg[\alpha_1(\vec{v}_i.\vec{E})+\alpha_2(\vec{v}_i.\vec{B})\nonumber\\
 & +\alpha_3 \vec{v}_i.(\vec{E}\times\vec{B})+\alpha_4(\vec{v}_i.\vec{\nabla}T)+\alpha_5\vec{v}_i.(\vec{\nabla}T\times \vec{B})+\nonumber\\&\alpha_6\vec{v}_i.(\vec{\nabla}T\times\vec{E})\bigg].
 \label{equnew25}
\end{align}
Comparing the coefficients of different tensor structures on both sides of Eq.\eqref{equnew25} we get,
\begin{align}
 & \alpha_6=0,\\
 & q_iE+\alpha_3 q_iB =-\frac{\omega_i}{\tau^{i}_R}\alpha_1,\label{equnew27}\\
 & \alpha_3 = \alpha_1 \tau^{i}_R \Omega_{c_i},\label{equnew28}\\
 & \alpha_2 = \alpha_3 \tau^{i}_R \Omega_{c_i}(\vec{E}.\vec{B})+b\tau^{i}_R\Omega_{c_i}(\vec{\nabla}T.\vec{B}),\label{equnew29}\\
 & \alpha_5= \Omega_{c_i}\tau^{i}_R\alpha_4,\label{equnew30}\\
 & \alpha_4 =\frac{\tau^{i}_R}{\omega_i}\bigg(\frac{\omega_i-b_ih}{T}\bigg)-\alpha_5 \tau^{i}_R\Omega_{c_i}\label{equnew31}.
\end{align}
where $\Omega_{c_i}=\frac{q_iB}{\omega_i}$ represents the cyclotron frequency of the particle with electric charge $q_i$. Using Eqs.\eqref{equnew27} to Eq.\eqref{equnew31}, we get
\begin{align}
 & \alpha_1 = -\frac{(q_iE)(\tau^{i}_R/\omega_i)}{1+(\Omega_{c_i}\tau^{i}_R)^2},\label{equnew33}\\
 & \alpha_4 = \frac{\tau^{i}_R}{\omega_i}\bigg(\frac{\omega_i- b_ih}{T}\bigg)\frac{1}{1+(\Omega_{c_i}\tau^{i}_R)^2}\label{equnew34}.
\end{align}
Using $\alpha_1$ and $\alpha_4$ as given in Eq.\eqref{equnew33} and Eq.\eqref{equnew34}, we can write
deviation from the equilibrium distribution function as,
\begin{align}
 \delta f_i &= \frac{\tau^{i}_R}{1+(\Omega_{c_i}\tau^{i}_R)^2} \bigg[q_i\bigg\{(\vec{v}_i.\vec{E})+(\Omega_{c_i}\tau^{i}_R)\vec{v}_i.(\vec{E}\times\vec{B})+\nonumber\\
& (\Omega_{c_i}\tau^{i}_R)^2(\vec{E}.\vec{B})(\vec{v}_i.\vec{B})\bigg\}
  -\bigg(\frac{\omega_i- b_ih}{T}\bigg)\bigg\{(\vec{v}_i.\vec{\nabla}T)+\nonumber\\
  &(\Omega_{c_i}\tau^{i}_R)\vec{v}_i.(\vec{\nabla T\times B})+(\Omega_{c_i}\tau^{i}_R)^2(\vec{\nabla}T.\vec{B})(\vec{v}_i.\vec{B})\bigg\}\bigg](-\frac{\partial f^{0}_{i}}{\partial\omega_i}).
 \label{equnew35}
 \end{align}
Now, we can express the electrical current and the heat current using $\delta f_i$ as given in Eq.\eqref{equnew35} as,
 \begin{align}
  j^l & =\sum_i g_i \int \frac{d^3k_i}{(2\pi)^3}q_i v_i^{l}\delta f_i\nonumber\\
  & = \sum_i \frac{g_iq_i}{3}\int\frac{d^3k_i}{(2\pi)^3}\frac{v_i^2 \tau^{i}_R}{1+(\Omega_{c_i}\tau^{i}_R)^2}\bigg[ q_i\delta^{lj}E^j+q_i (\Omega_{c_i}\tau^{i}_R)\nonumber\\
  &\epsilon^{ljk}h^kE^j+q_i(\Omega_{c_i}\tau^{i}_R)^2h^lh^jE^j
   -\bigg(\frac{\omega_i- b_ih}{T}\bigg)\bigg\{\delta^{lj}\frac{\partial T}{\partial x^j}\nonumber\\
  & +(\Omega_{c_i}\tau^{i}_R)\epsilon^{ljk}h^k \frac{\partial T}{\partial x^j}+(\Omega_{c_i}\tau^{i}_R)^2h^lh^j\frac{\partial T}{\partial x^j}\bigg\}\bigg](-\frac{\partial f_i^{0}}{\partial \omega_i}),
  \label{equnew36}
 \end{align}
and,
\begin{align}
 {I}^l &  = \sum_i g_i \int \frac{d^3k_i}{(2\pi)^3}v_i^l\bigg(\omega_i-b_ih\bigg)\delta f_i\nonumber\\
 & = \sum_i \frac{g_i}{3}\int\frac{d^3k_i}{(2\pi)^3}\frac{v_i^2\tau^{i}_R}{1+(\Omega_{c_i}\tau^{i}_R)^2}\bigg(\omega_i-b_ih\bigg)\bigg[ q_i\delta^{lj}E^j \nonumber\\
 &+q_i (\Omega_{c_i}\tau^{i}_R)\epsilon^{ljk}h^kE^j+q_i(\Omega_{c_i}\tau^{i}_R)^2h^lh^jE^j
  -\bigg(\frac{\omega_i- b_ih}{T}\bigg)\nonumber\\
  &\bigg\{\delta^{lj}\frac{\partial T}{\partial x^j}+(\Omega_{c_i}\tau^{i}_R)\epsilon^{ljk}h^k \frac{\partial T}{\partial x^j}+(\Omega_{c_i}\tau^{i}_R)^2h^lh^j\frac{\partial T}{\partial x^j}\bigg\}\bigg]\nonumber\\
  &~~~~~~~~~~~~~~~~~~~~~~~~~~~~~~~~~~~~~~~~~~~~~~~~~~~~~~~~~(-\frac{\partial f_i^{0}}{\partial \omega_i}).
  \label{equnew37}
\end{align}
Here, to simplify the calculation,  we can choose the magnetic field along the $z$ direction. The direction of the electric field and the temperature gradient are perpendicular to the $z$ axis {\it i.e.} it is in the $x-y$ plane. Under these conditions, the components of the electric current in the $x-y$ plane are given as,
\begin{align}
 j_x = &  \sum_i\frac{g_iq_i}{3}\int\frac{d^3k_i}{(2\pi)^3}\frac{v_i^2q_i\tau^{i}_R}{1+(\Omega_{c_i}\tau^{i}_R)^2}\bigg[E_x+(\Omega_{c_i}\tau^{i}_R)E_y\bigg]\nonumber\\
 &(-)\frac{\partial f_i^{0}}{\partial \omega_i}
  -\sum_i\frac{g_i q_i}{3T}\int\frac{d^3k_i}{(2\pi)^3}\frac{v_i^2\tau^{i}_R\left(\omega_i-b_ih\right)}{1+(\Omega_{c_i}\tau)^2}\nonumber\\
 & \bigg[\frac{dT}{dx}+(\Omega_{c_i}\tau^{i}_R)\frac{dT}{dy}\bigg]
 (-\frac{\partial f_i^{0}}{\partial \omega_i}),
 \label{equnew38}
\end{align}
and,
\begin{align}
 j_y = &  \sum_i\frac{g_iq_i}{3}\int\frac{d^3k_i}{(2\pi)^3}\frac{v_i^2q_i\tau^{i}_R}{1+(\Omega_{c_i}\tau^{i}_R)^2}\bigg[E_y-(\Omega_{c_i}\tau^{i}_R)E_x\bigg]\nonumber\\
 &(-)\frac{\partial f_i^{0}}{\partial \omega_i}
  -\sum_i\frac{g_i q_i}{3T}\int\frac{d^3k_i}{(2\pi)^3}\frac{v_i^2\tau^{i}_R\left(\omega_i-b_ih\right)}{1+(\Omega_{c_i}\tau)^2}\nonumber\\
  &\bigg[\frac{dT}{dy}-(\Omega_{c_i}\tau^{i}_R)\frac{dT}{dx}\bigg](-\frac{\partial f_i^{0}}{\partial \omega_i}).
 \label{equnew39}
\end{align}
Equations \eqref{equnew38} and \eqref{equnew39} can be written in a compact form by introducing the following integrals,
\begin{align}
 & H_{1_i}=\frac{g_i}{3}\int\frac{d^3k_i}{(2\pi)^3}\frac{\tau^{i}_R}{1+(\Omega_{c_i}\tau^{i}_R)^2}\left(\frac{\vec{k_i}^2}{\omega_i^2}\right)(-\frac{\partial f_i^{0}}{\partial \omega_i}),\label{equnew40}\\
& H_{2_i}=\frac{g_i}{3}\int\frac{d^3k_i}{(2\pi)^3}\frac{\tau^{i}_R(\Omega_{c_i}\tau^{i}_R)}{1+(\Omega_{c_i}\tau^{i}_R)^2}\left(\frac{\vec{k_i}^2}{\omega_i^2}\right)(-\frac{\partial f_i^{0}}{\partial \omega_i}),\label{equnew41}\\
& H_{3_i}=\frac{g_i}{3}\int\frac{d^3k_i}{(2\pi)^3}\frac{\tau^{i}_R\omega_i}{1+(\Omega_{c_i}\tau^{i}_R)^2}\left(\frac{\vec{k_i}^2}{\omega_i^2}\right)(-\frac{\partial f_i^{0}}{\partial \omega_i}),\label{equnew42}\\
& H_{4_i}=\frac{g_i}{3}\int\frac{d^3k_i}{(2\pi)^3}\frac{\tau^{i}_R\omega_i(\Omega_{c_i}\tau^{i}_R)}{1+(\Omega_{c_i}\tau^{i}_R)^2}\left(\frac{\vec{k_i}^2}{\omega_i^2}\right)(-\frac{\partial f_i^{0}}{\partial \omega_i}).\label{equnew43}
 \end{align}
The integrals as given in Eqs.\eqref{equnew40}-\eqref{equnew43} allows us to write  Eq.\eqref{equnew38} and  Eq.\eqref{equnew39}, respectively, as 
\begin{widetext}
\begin{align}
 j_x = \sum_i q_i^2H_{1_i}E_x+\sum_i q_i^2H_{2_i}E_y -\frac{1}{T}\sum_{a}q_i\bigg(H_{3_i}-b_ihH_{1_i}\bigg)\frac{d T}{dx}-\frac{1}{T}\sum_{a}q_i\bigg(H_{4_i}-b_ihH_{2_i}\bigg)\frac{d T}{dy},
 \label{equnew44}
\end{align}
and,
\begin{align}
 j_y = \sum_i q_i^2H_{1_i}E_y-\sum_i q_i^2H_{2_i}E_x -\frac{1}{T}\sum_{a}q_i\bigg(H_{3_i}-b_ihH_{1_i}\bigg)\frac{d T}{dy}+\frac{1}{T}\sum_{a}q_i\bigg(H_{4_i}-b_ihH_{2_i}\bigg)\frac{d T}{dx}.
 \label{equnew45}
\end{align}
Here, in the presence of a magnetic field, the magneto-Seebeck coefficient ($S_B$) can be determined by setting $j_x = j_y=0$ so that the electric field becomes proportional to the temperature gradient. For $j_x=0$ and $j_y=0$ we can solve Eqs.\eqref{equnew44} and \eqref{equnew45} to get $E_x$ and $E_y$ in terms of temperature gradients $\frac{dT}{dx}$ and  $\frac{dT}{dy}$ as,
\begin{align}
 E_x &  = \frac{\sum_i q_i^2H_{1_i}\sum_iq_i(H_{3_i}-b_ihH_{1_i})+\sum_i q_i^2H_{2_i}\sum_iq_i(H_{4_i}-b_ihH_{2_i})}{T\bigg[\bigg(\sum_i q_i^2H_{1_i}\bigg)^2+\bigg(\sum_i q_i^2H_{2_i}\bigg)^2\bigg]}\frac{dT}{dx}\nonumber\\
 & +\frac{\sum_i q_i^2H_{1_i}\sum_iq_i(H_{4_i}-b_ihH_{2_i})-\sum_i q_i^2H_{2_i}\sum_iq_i(H_{3_i}-b_ihH_{1_i})}{T\bigg[\bigg(\sum_i q_i^2H_{1_i}\bigg)^2+\bigg(\sum_i q_i^2H_{2_i}\bigg)^2\bigg]}\frac{dT}{dy},
 \label{equnew46}
\end{align}
and,
\begin{align}
 E_y &  = \frac{\sum_i q_i^2H_{2_i}\sum_iq_i(H_{3_i}-b_ihH_{1_i})-\sum_i q_i^2H_{1_i}\sum_iq_i(H_{4_i}-b_ihH_{2_i})}{T\bigg[\bigg(\sum_i q_i^2H_{1_i}\bigg)^2+\bigg(\sum_i q_i^2H_{2_i}\bigg)^2\bigg]}\frac{dT}{dx}\nonumber\\
 & +\frac{\sum_i q_i^2H_{1_i}\sum_iq_i(H_{3_i}-b_ihH_{1_i})+\sum_i q_i^2H_{2_i}\sum_iq_i(H_{4_i}-b_ihH_{2_i})}{T\bigg[\bigg(\sum_i q_i^2H_{1_i}\bigg)^2+\bigg(\sum_i q_i^2H_{2_i}\bigg)^2\bigg]}\frac{dT}{dy}.
 \label{equnew47}
\end{align}
\end{widetext}
Equations \eqref{equnew46} and \eqref{equnew47} can be written in a compact form in the following way,
\begin{align}
 \begin{pmatrix}
E_x \\
\\
E_y 
\end{pmatrix}= \begin{pmatrix}
S_B & NB \\
\\
-NB & S_B 
\end{pmatrix}\begin{pmatrix}
\frac{dT}{dx} \\
\\
\frac{dT}{dy} 
\end{pmatrix},
\end{align}
here one can identify the magneto-Seebeck coefficient as, 
\begin{widetext}
\begin{align}
 S_B & = \frac{\sum_i q_i^2H_{1_i}\sum_iq_i(H_{3_i}-b_ihH_{1_i})+\sum_i q_i^2H_{2_i}\sum_iq_i(H_{4_i}-b_ihH_{2_i})}{T\bigg[\bigg(\sum_i q_i^2H_{1_i}\bigg)^2+\bigg(\sum_i q_i^2H_{2_i}\bigg)^2\bigg]}\nonumber\\
 & = \frac{(\sigma_{el}/T)(\mathcal{I}_{31}/T^2)+(\sigma_{H}/T)(\mathcal{I}_{42}/T^2)}{(\sigma_{el}/T)^2+(\sigma_{H}/T)^2},
 \label{equnew49}
\end{align}
and the normalized Nernst coefficient ($NB$) is given as,
\begin{align}
 NB & = \frac{\sum_i q_i^2H_{1_i}\sum_iq_i(H_{4_i}-b_ihH_{2_i})-\sum_i q_i^2H_{2_i}\sum_iq_i(H_{3_i}-b_ihH_{1_i})}{T\bigg[\bigg(\sum_i q_i^2H_{1_i}\bigg)^2+\bigg(\sum_i q_i^2H_{2_i}\bigg)^2\bigg]}\nonumber\\
 & = \frac{(\sigma_{el}/T)(\mathcal{I}_{42}/T^2)-(\sigma_{H}/T)(\mathcal{I}_{31}/T^2)}{(\sigma_{el}/T)^2+(\sigma_{H}/T)^2}.
 \label{equnew50}
 \end{align}
 \end{widetext}
Here, we have identified the electrical conductivity
in the presence of a magnetic field and the Hall conductivity as $\sigma_{el}=\sum_iq_i^2H_{1_i}$ and $\sigma_{H}=\sum_iq_i^2H_{2_i}$ respectively \cite{Singh:2024emy}. The integrals $\mathcal{I}_{31}$ and $\mathcal{I}_{42}$ in Eqs.\eqref{equnew49} and \eqref{equnew50} are defined as  $\mathcal{I}_{31} = \sum_iq_i(H_{3_i}-b_ihH_{1_i})$ and $\mathcal{I}_{42}\equiv \sum_iq_i(H_{4_i}-b_ihH_{2_i})$.
Note that in the absence of a magnetic field, integrals $H_{2_i}$ and $H_{4_i}$ are identically zero. Hence, the normalized Nernst coefficient vanishes in the absence of a magnetic field, and the magneto-Seebeck coefficient turns into the Seebeck coefficient in the absence of a magnetic field as given in Eq.\eqref{equnew11}. 

In the above calculation, for the sake of simplicity, we take the thermal averaged relaxation time after integrating energy-dependent relaxation time over the equilibrium distribution function. The thermal averaged relaxation time ($\tau_{R}^i$) for $i$th hadron species can be expressed in terms of scattering cross section as~\cite{Das:2020beh},
\begin{align}
    {\tau_{R}^i}^{-1} = \sum_j n_j \langle\sigma_{ij}v_{ij}\rangle
\end{align}
where,
\begin{align}
 \langle\sigma_{ij}v_{ij}\rangle &= \frac{\sigma}{8Tm_i^2m_j^2\mathcal{K}_2(m_i/T)\mathcal{K}_2(m_j/T)}~ \times\nn\\
 &\int_{(m_i+m_j)^2}^{\infty}ds~\times \nn\\
 &\frac{[s-(m_i-m_j)^2]}{\sqrt{s}}
 \times [s-(m_i+m_j)^2]\mathcal{K}_1(\sqrt{s}/T),
 \label{equnew70}
\end{align}
here, $\sigma = 4\pi r_h^2$ is the total scattering cross section for the hard spheres, and it is independent of both temperature and baryon chemical potential. $\mathcal{K}_1, \mathcal{K}_2$ are modified Bessel functions of the first and second order.

\section{Results and Discussions}\label{sec-result}
We take the different versions of hadron resonance gas models with discrete particle spectrum, including all the hadrons and their resonances up to the mass cutoff $\Lambda$ = 2.6 GeV~\cite{ParticleDataGroup:2008zun}. The hard-scattering approximation is used to estimate the relaxation time of the hadrons. Here, we consider a uniform radius, $r_m$ = 0.2 fm for mesons and  $r_{b,\bar{b}}$ = 0.62 fm for baryons and antibaryons~\cite{Pradhan:2022gbm}. For the VDWHRG model, the van der Waals parameters are taken as $a$ = 0.926 GeV fm$^3$ and $b$ = (16/3)$\pi r^{3}$~\cite{Pradhan:2022gbm, Samanta:2017yhh}. In this section, we present the results for different thermoelectric coefficients of a hot and dense hadron gas both in the absence and presence of a nonzero external magnetic field. 
\label{results}
\subsection{In the absence of magnetic field}
\begin{figure*}
	\centering
	\includegraphics[scale=0.27]{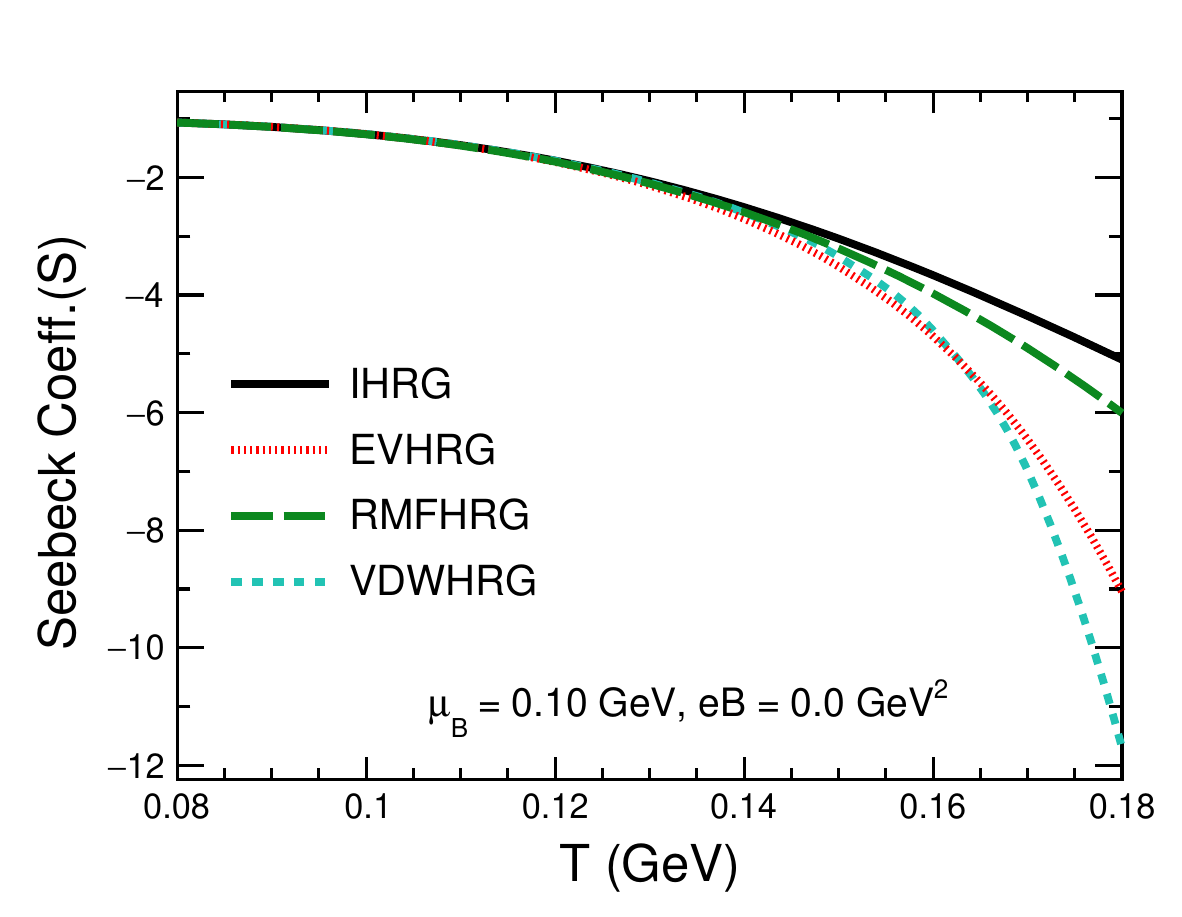}
 \includegraphics[scale=0.27]{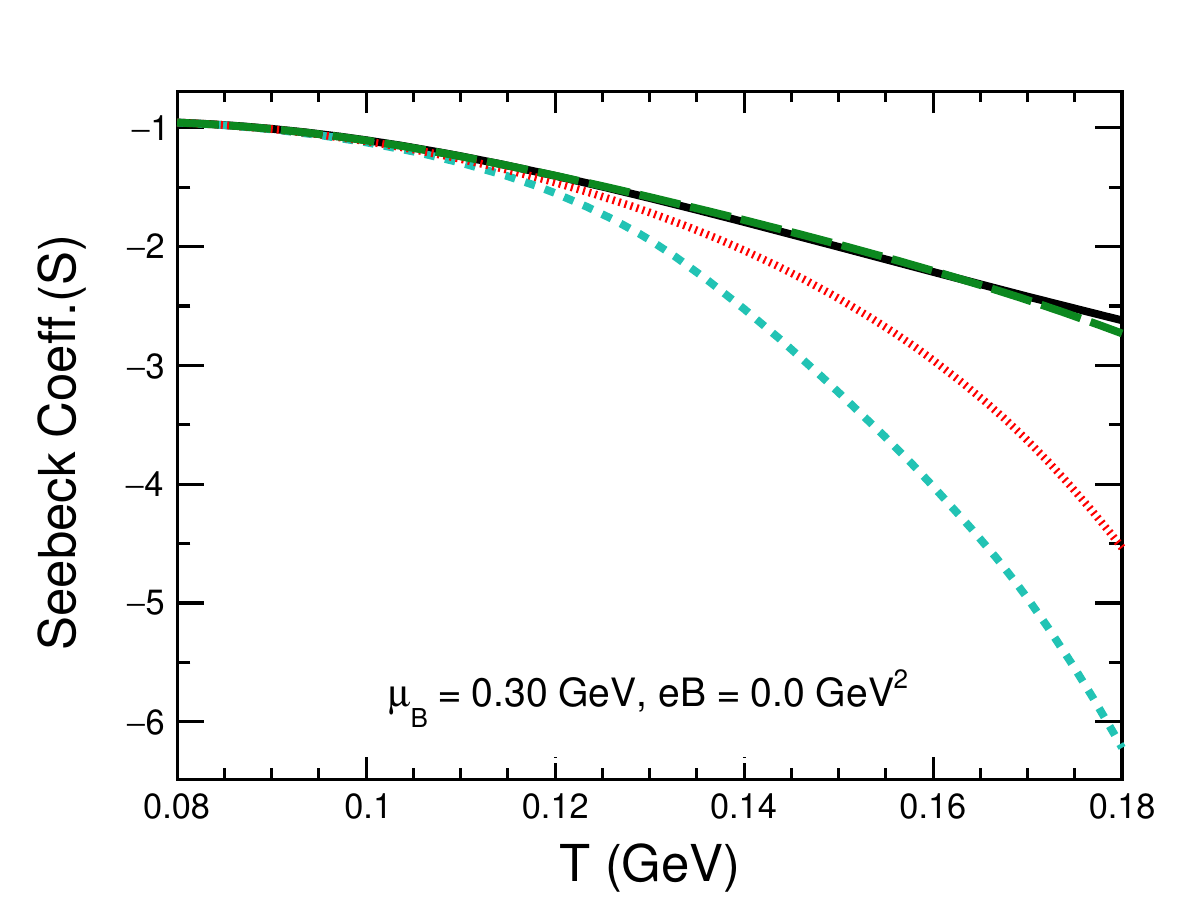}
  \includegraphics[scale=0.27]{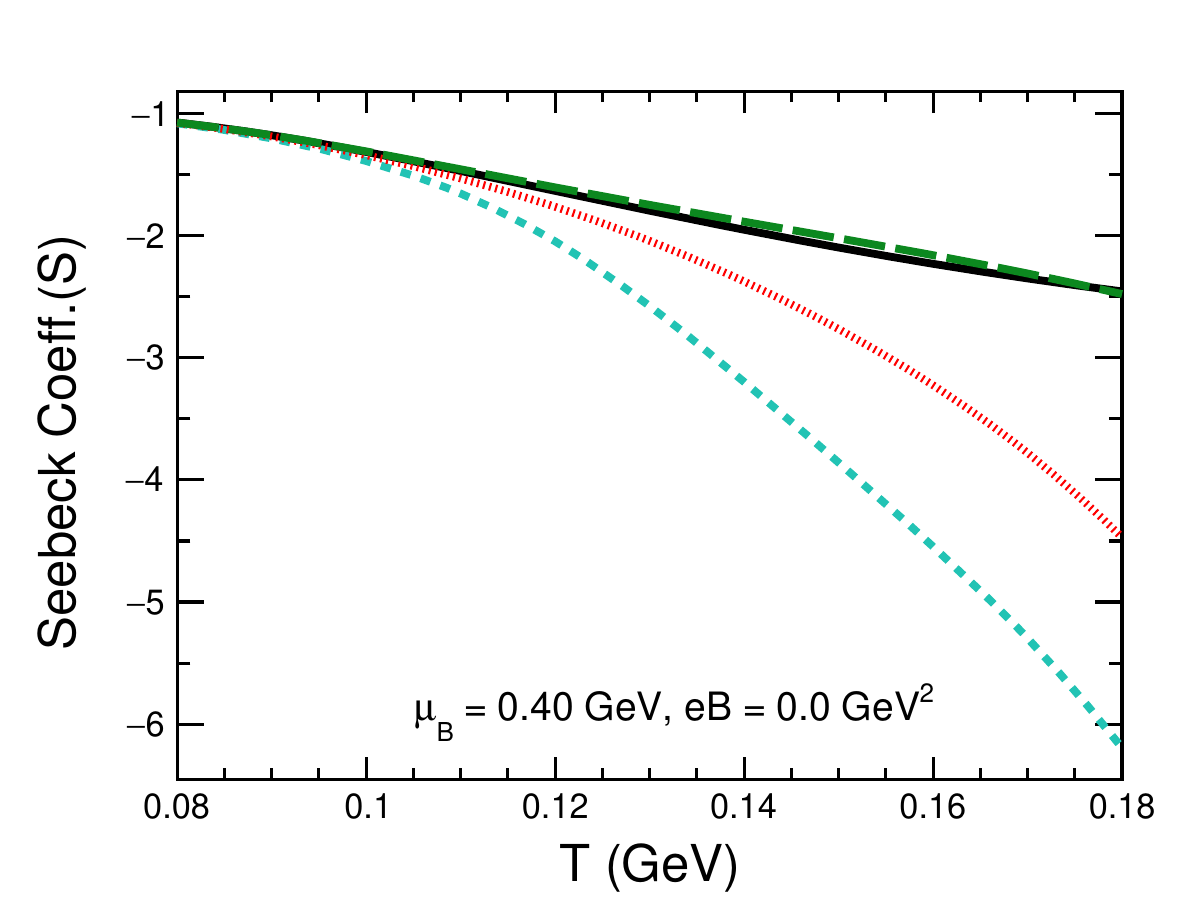}
	\caption{Seebeck coefficient ($S$) obtained in different hadronic models as a function of temperature at $\mu_{B}$ = 0.10 GeV (left panel), 0.30 GeV (middle panel), and 0.40 GeV (right panel).}
	\label{Fig-seebeck1}
\end{figure*}

\begin{figure}
	\centering
	\includegraphics[scale=0.40]{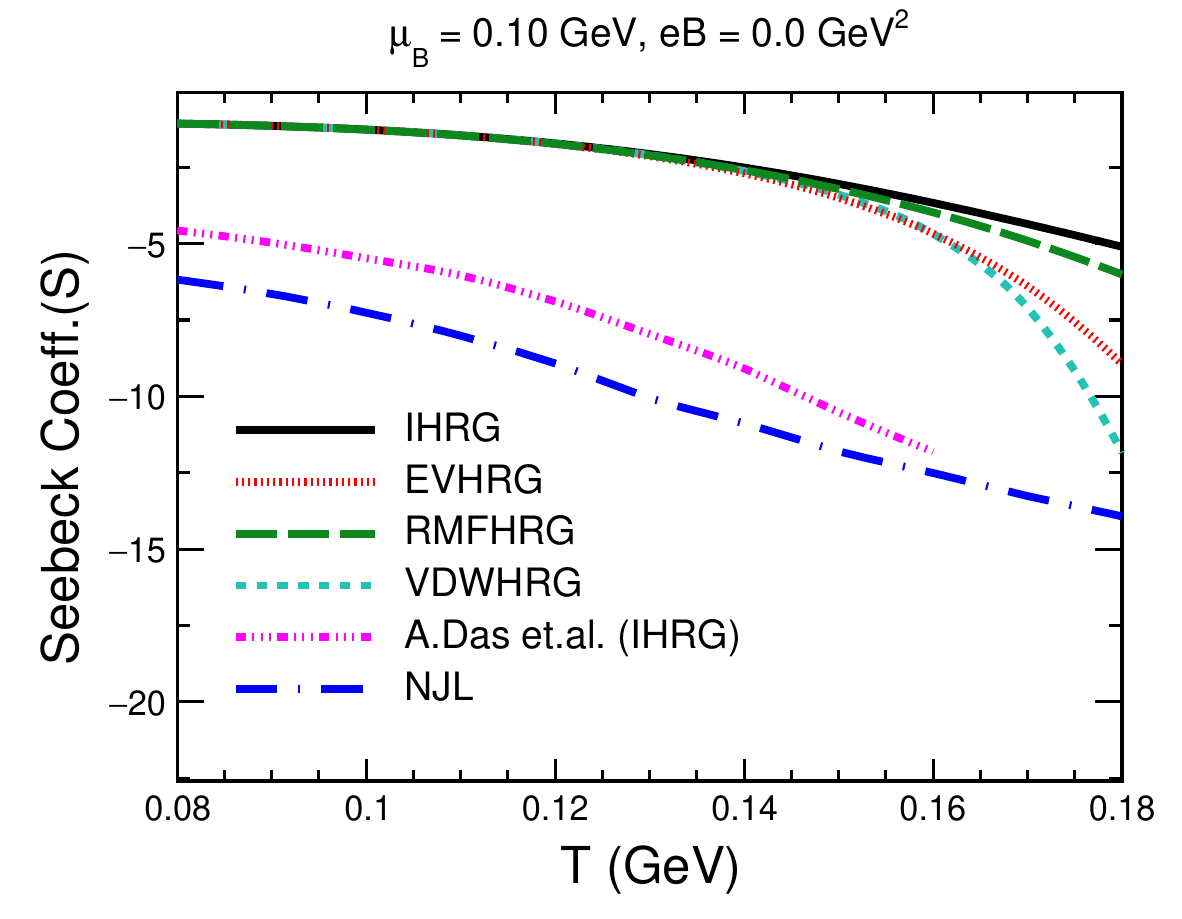}
	\caption{Comparison of Seebeck coefficient ($S$) in different models as a function of temperature at a baryonic chemical potential $\mu_{B}$ =  0.10 GeV. The dotted magenta line shows the results obtained in ideal HRG in Ref. \cite{Das:2020beh}, and the dashed blue line is obtained in the Nambu–Jona Lasinio (NJL) model \cite{Abhisek2022}.}
	\label{Fig-seecom}
\end{figure}

\begin{figure*}
	\centering
	\includegraphics[scale=0.27]{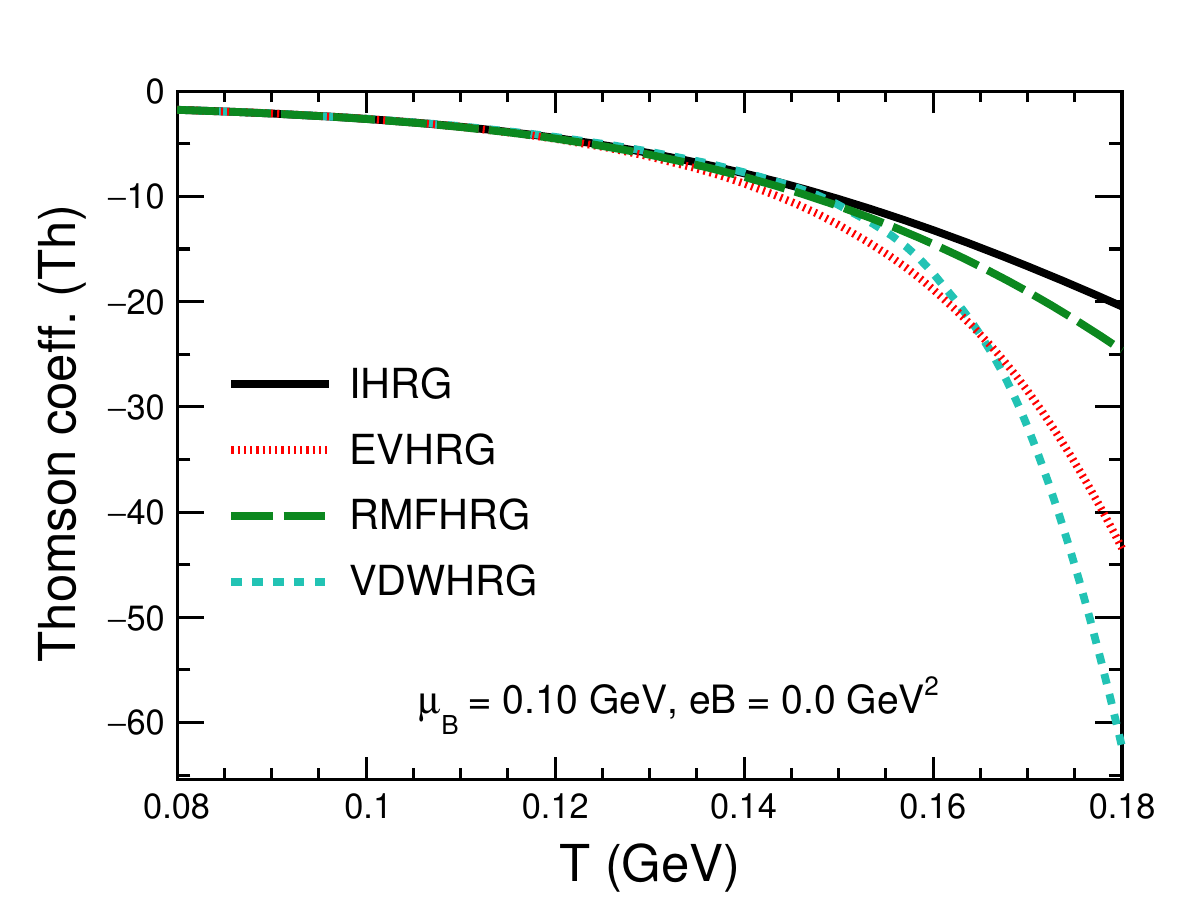}
 \includegraphics[scale=0.27]{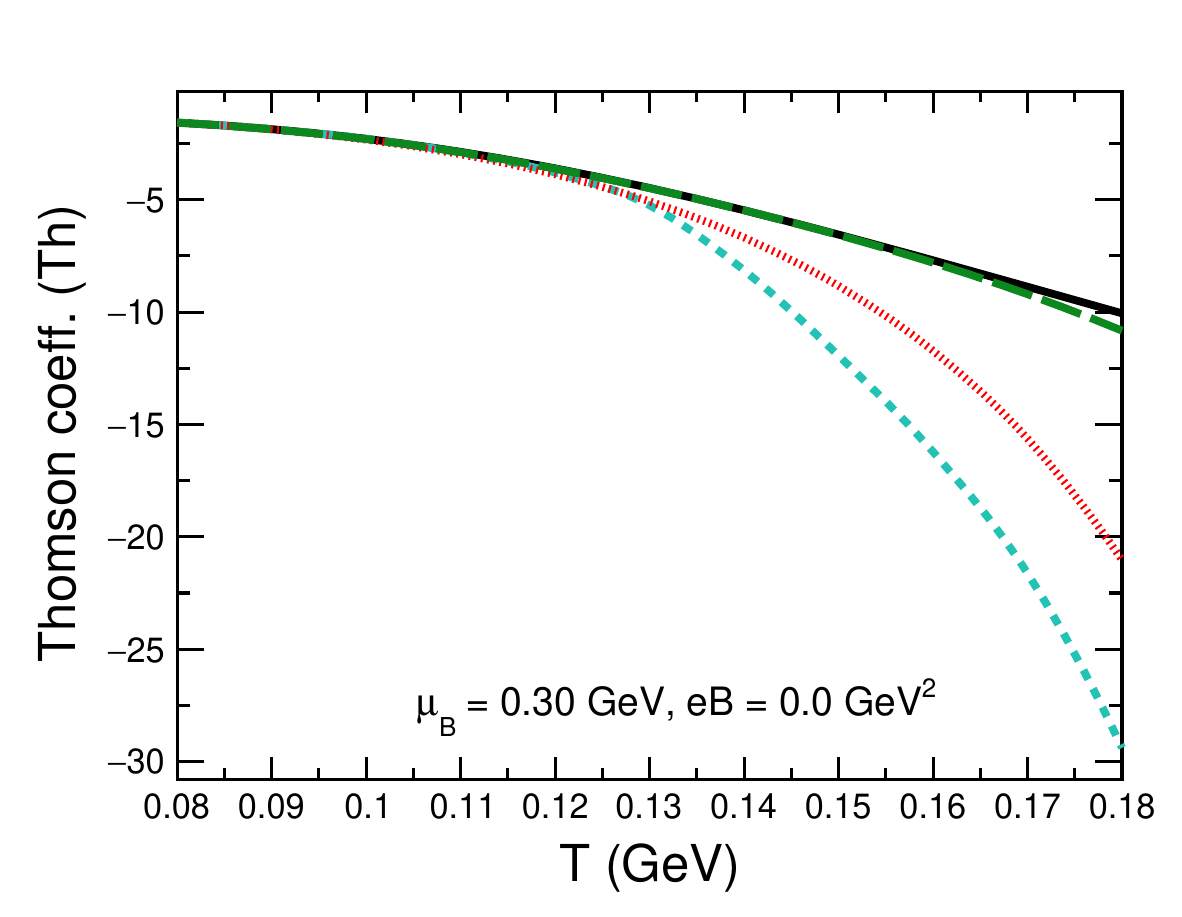}
  \includegraphics[scale=0.27]{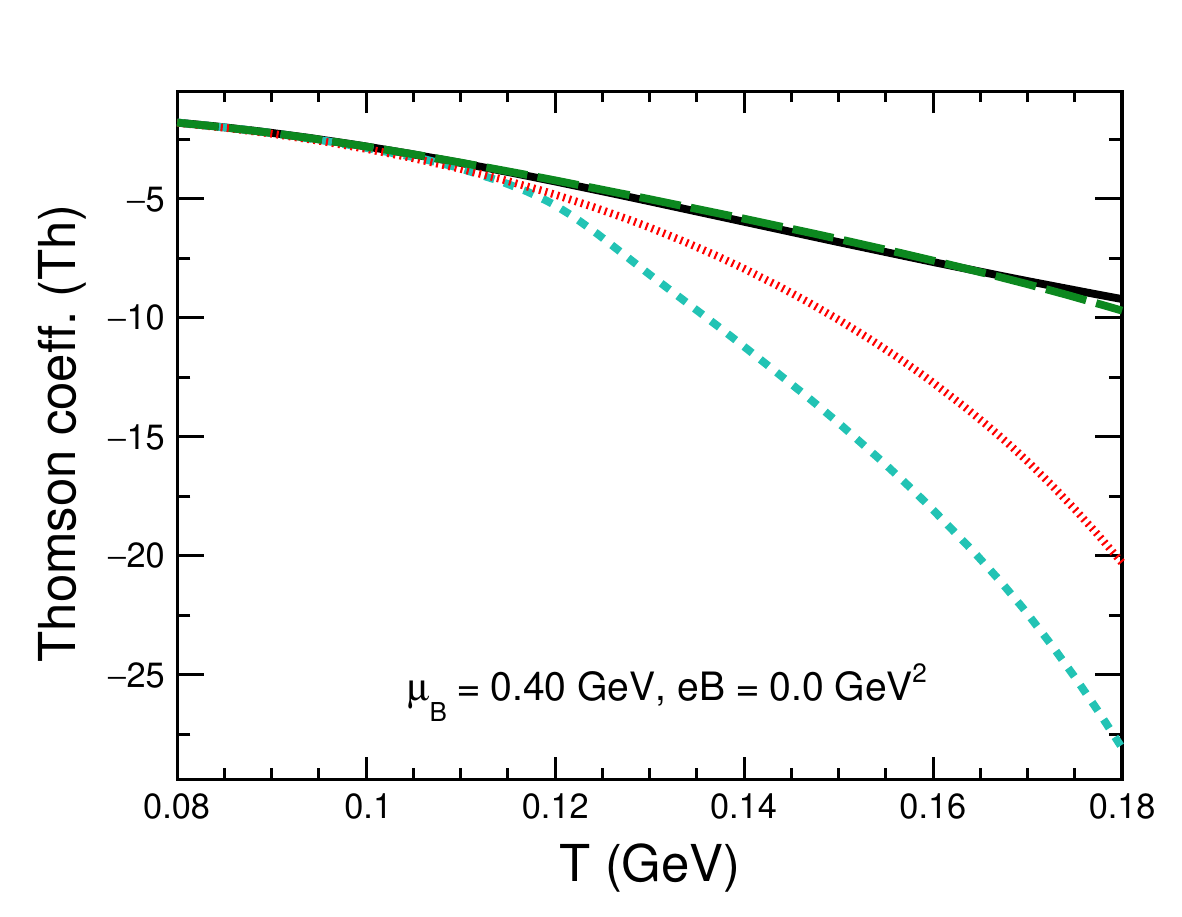}
	\caption{Thomson coefficient ($Th$) obtained in different hadronic models as a function of temperature at $\mu_{B}$ = 0.10 GeV (left panel), 0.30 GeV (middle panel), and 0.40 GeV (right panel).}
	\label{Fig-Thomson1}
\end{figure*}

\begin{figure*}
	\centering
	\includegraphics[scale=0.27]{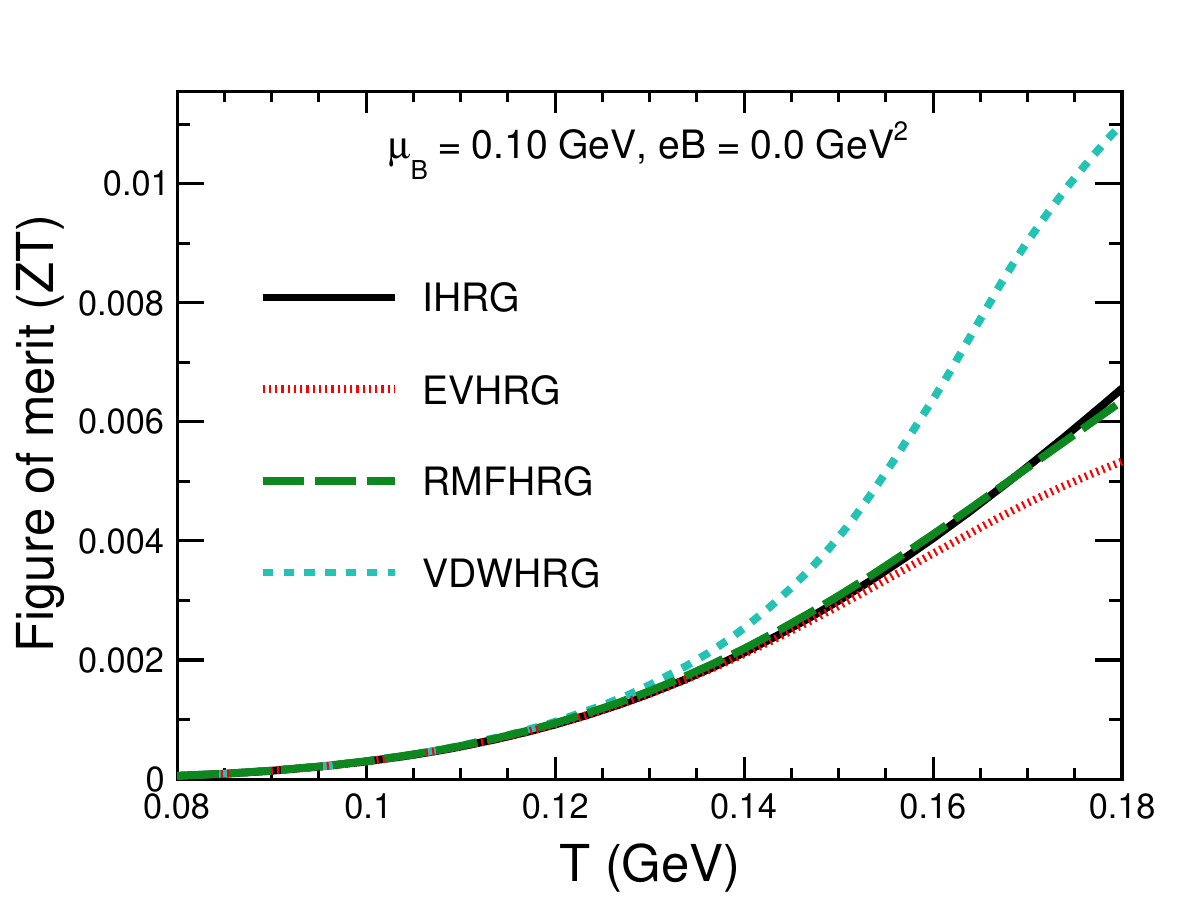}
 \includegraphics[scale=0.27]{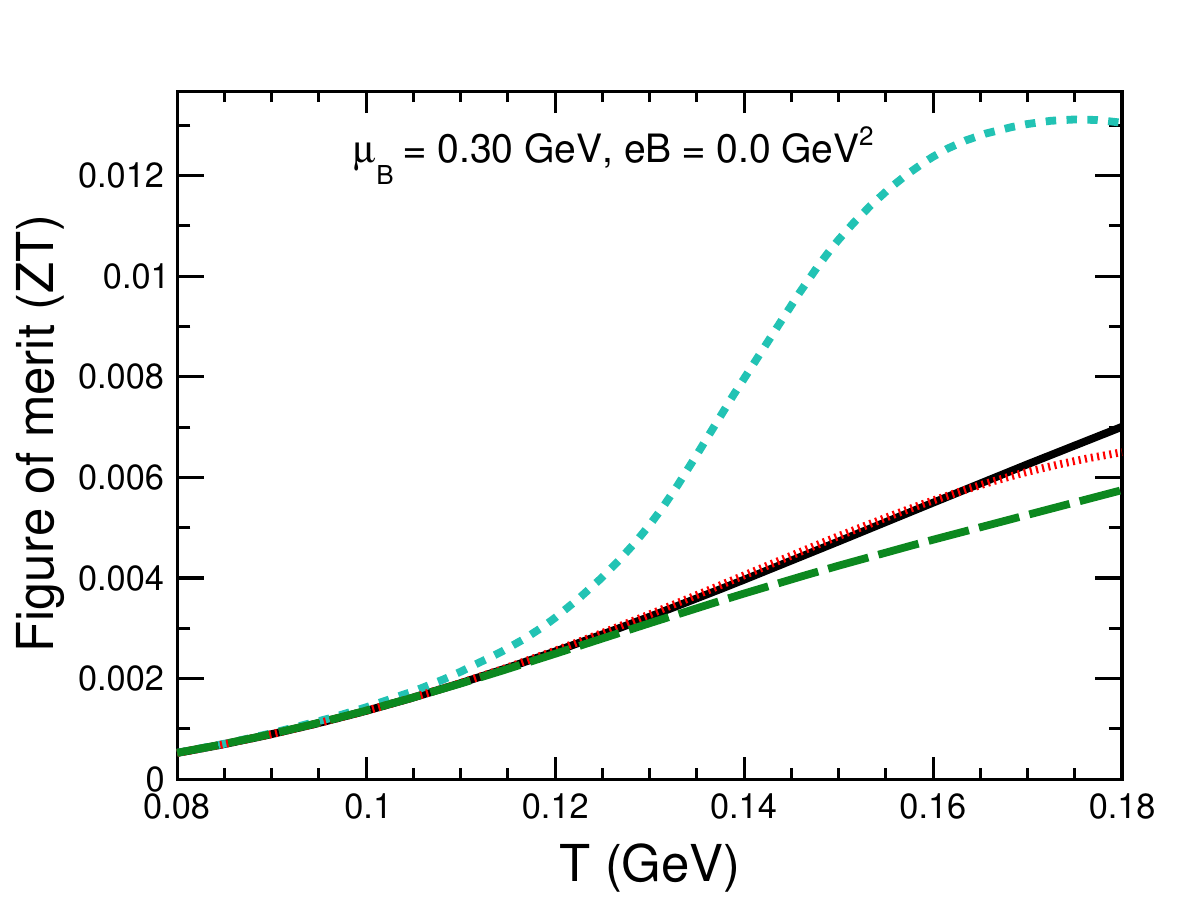}
  \includegraphics[scale=0.27]{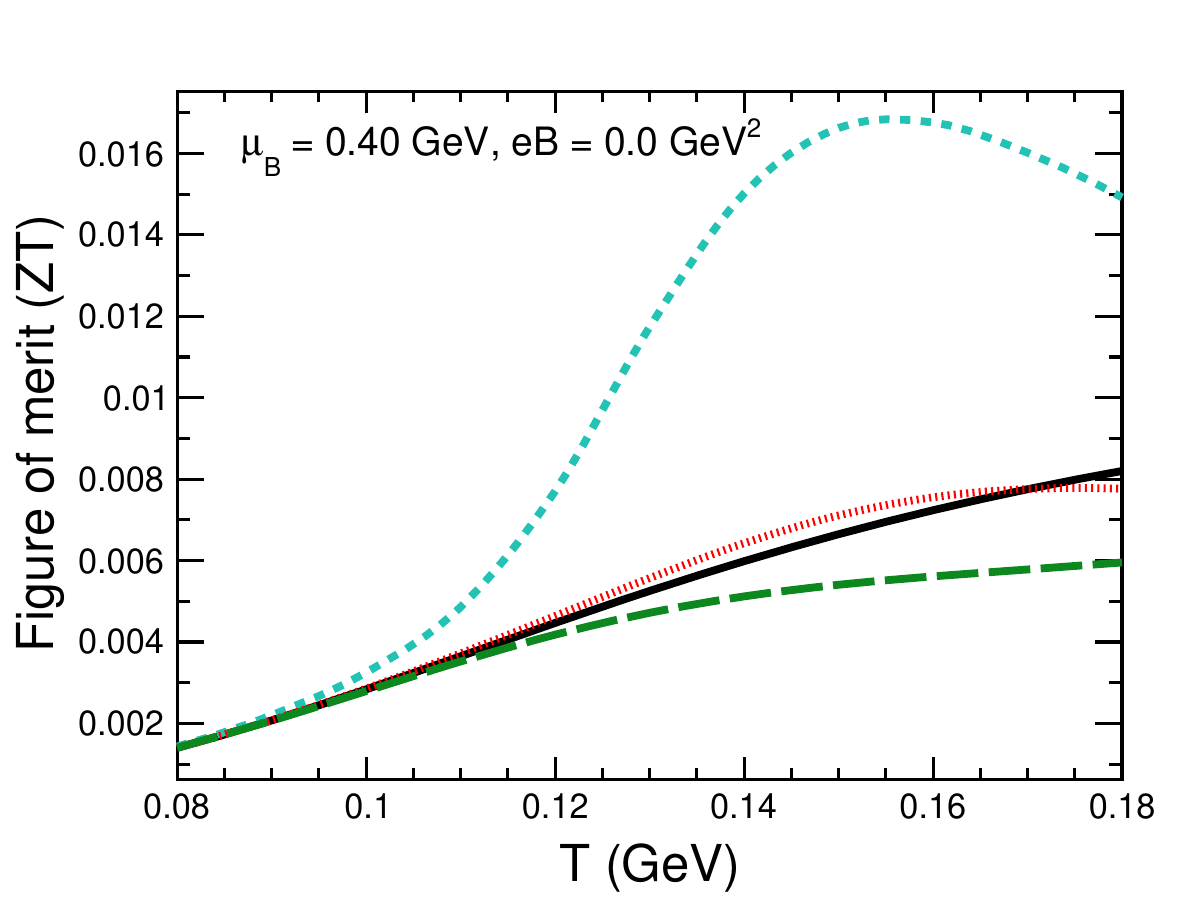}
	\caption{Thermoelectric figure of merit ($ZT$) as a function of temperature at magnetic field $eB =  0.0~GeV^{2}$ and baryonic chemical potential at $\mu_{B}$  left: 0.10 GeV, middle: 0.30 GeV, right:  0.40 GeV.}
	\label{Fig-ZT}
\end{figure*}
Figure~\ref{Fig-seebeck1} shows the Seebeck coefficient, $S$ as a function of temperature for three different baryon chemical potentials at  0.10, 0.30, and 0.40 GeV, in the absence of an external magnetic field. The Seebeck coefficient obtained in the IHRG model is shown by a black solid line, whereas the red dotted line represents the EVHRG model. The green and cyan dashed lines represent the results for the RMFHRG and VDWHRG models, respectively. Here, we observe that for the given range of temperature, $S$ is negative for a hot hadron gas, and is universally decreasing with an increase in temperature. However, the degree of decrease depends on the model. The heat flow for a relativistic system can only be defined for any conserved current. For the case of hadron gas, the heat current corresponds to the net baryon current. It is to be noted that baryons have a dominant contribution to the Seebeck coefficient, whereas the contribution of mesons comes only through the enthalpy of the system. For the case of HRG models, the entropy production for lighter mesons, such as pions and kaons, is substantial. Therefore, the enthalpy per baryon ($h$) exceeds the single particle energy ($\omega_{i}$), hence the term ($\omega_{i} - b_{i}h$) in $\mathcal{I}_{1}/T^2$ becomes negative in Eq.~(\ref{equnew11}), and as a result, $S$ becomes negative. The negative sign of $S$ indicates that the induced electric field in the medium is aligned opposite to the temperature gradient. For the condensed matter systems, $S$ can be both positive or negative corresponding to the contribution of holes and electrons, respectively. In our case, we observe that the scaled electrical conductivity, $\sigma_{el}/T$, decreases as a function of temperature, in line with what was observed in earlier studies \cite{Das:2020beh, Pradhan:2022gbm, Singh:2023pwf, Das:2019wjg}. Therefore, the magnitude of the Seebeck coefficient increases as a function of temperature but in a negative direction. As shown in the left panel of Fig.~\ref{Fig-seebeck1}, results from all the models coincide up to a temperature $T\approx 0.14$ GeV, after which deviations are observed. The effect of interactions between the hadrons seems to be prominent after this temperature. One can observe from Fig.~\ref{Fig-seebeck1} that as $\mu_{B}$ increases, $S$ increases for all the models and the deviation from each model starts appearing even at low temperatures; around $T\approx 0.12$ GeV for $\mu_B=$ 0.30 GeV and $T\approx 0.11$ GeV for $\mu_B=$ 0.40 GeV.

\begin{figure*}
	\centering
	\includegraphics[scale=0.27]{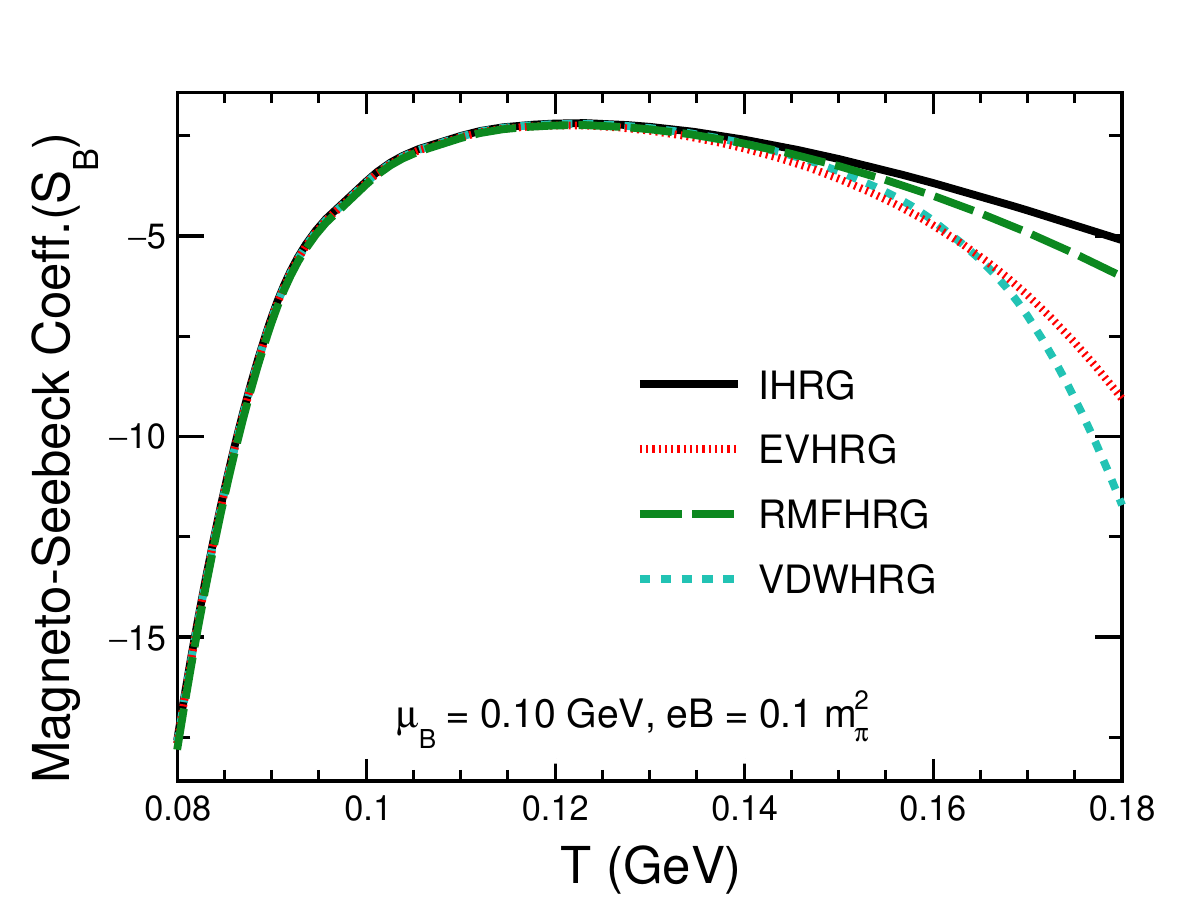}
    \includegraphics[scale=0.27]{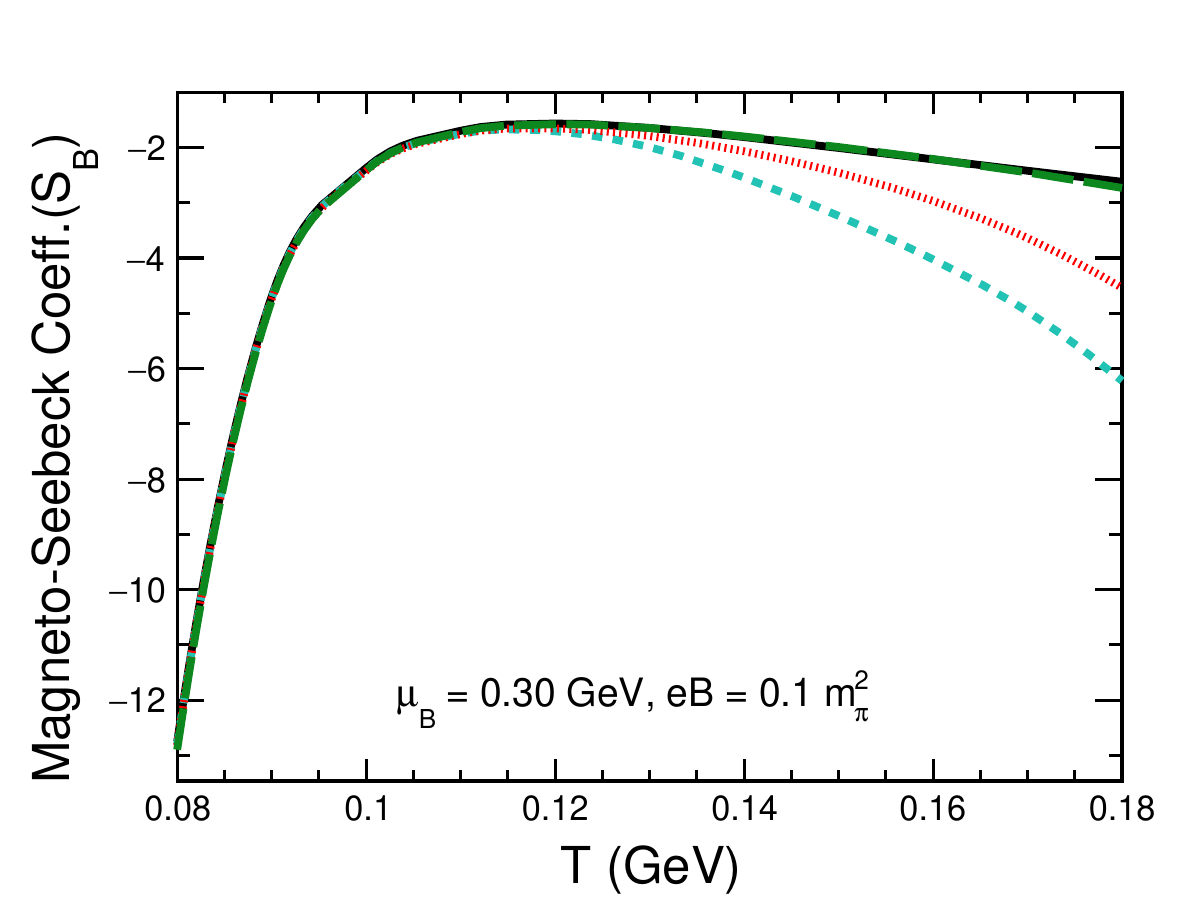}
    \includegraphics[scale=0.27]{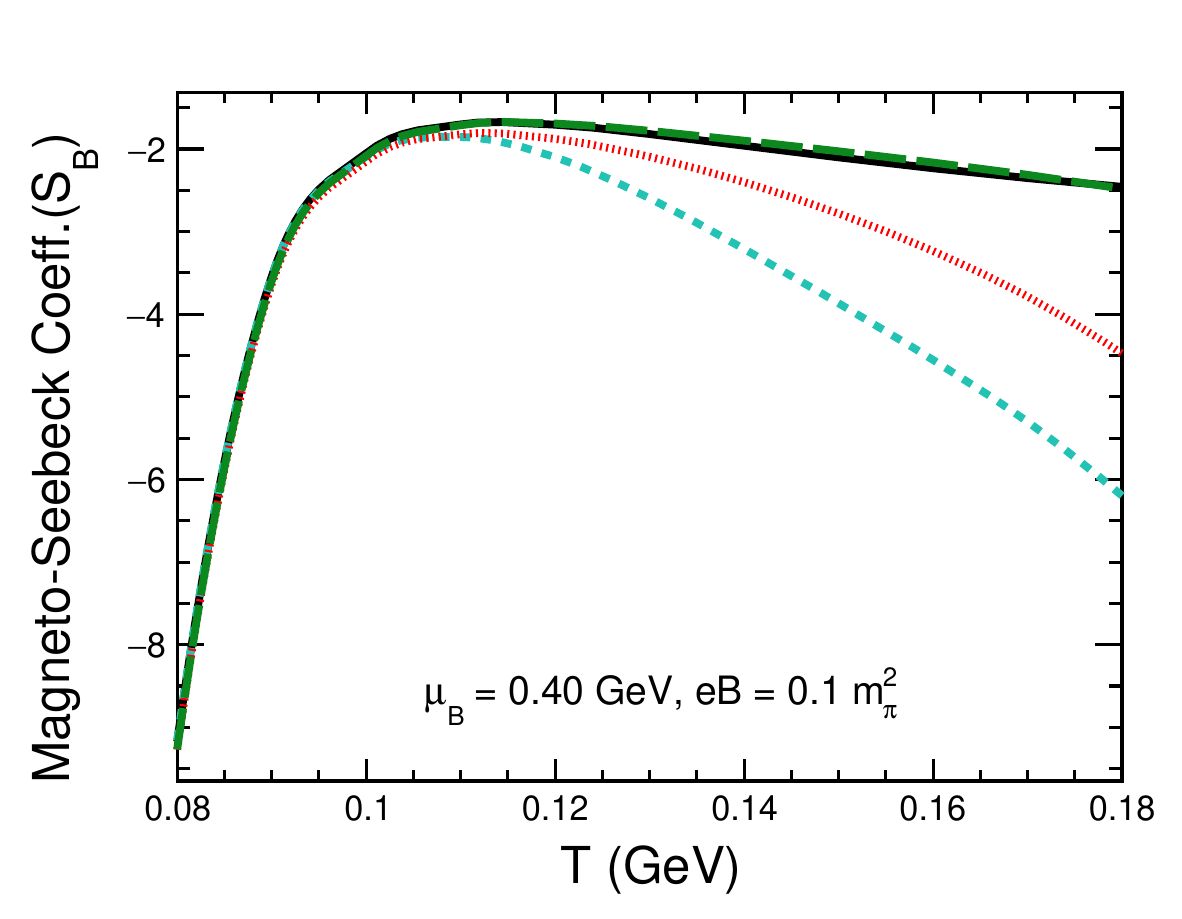}

	\includegraphics[scale=0.27]{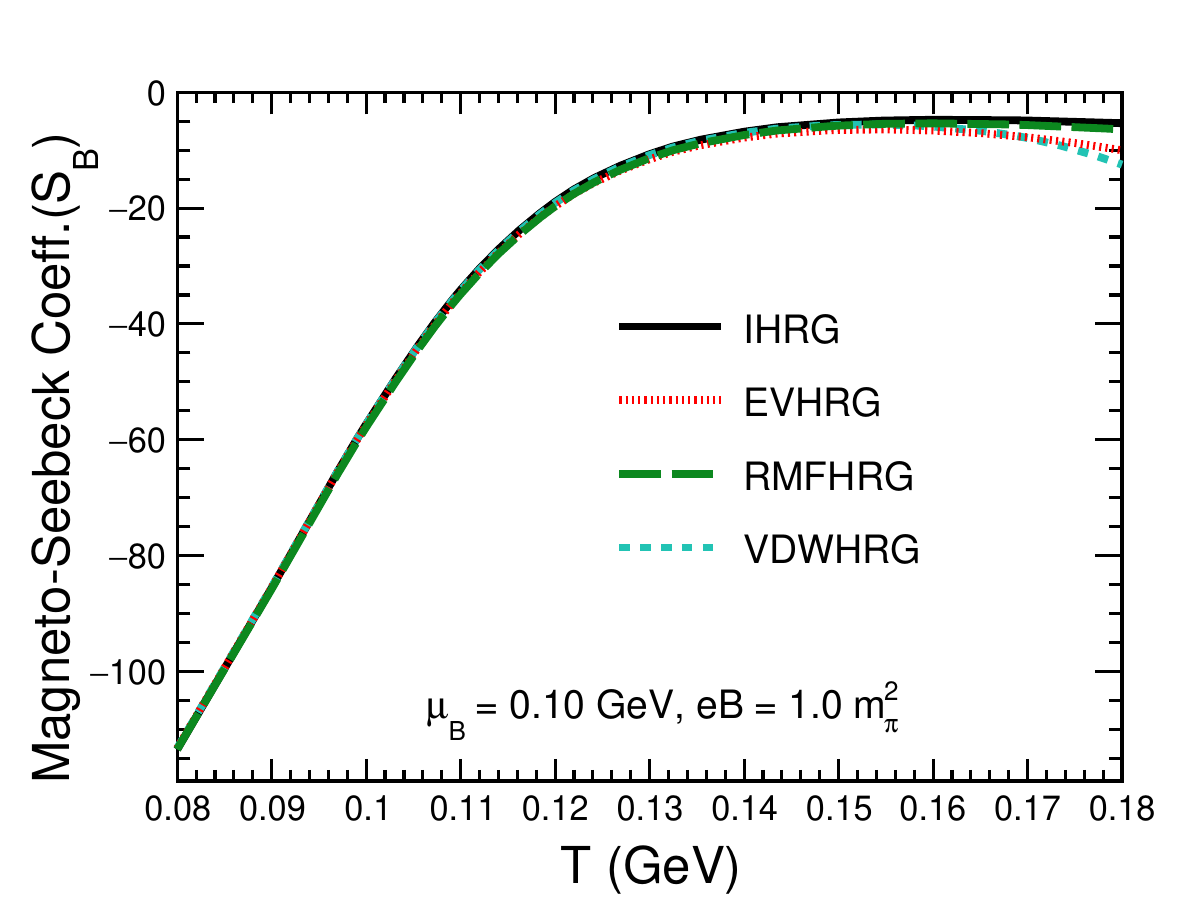}
    \includegraphics[scale=0.27]{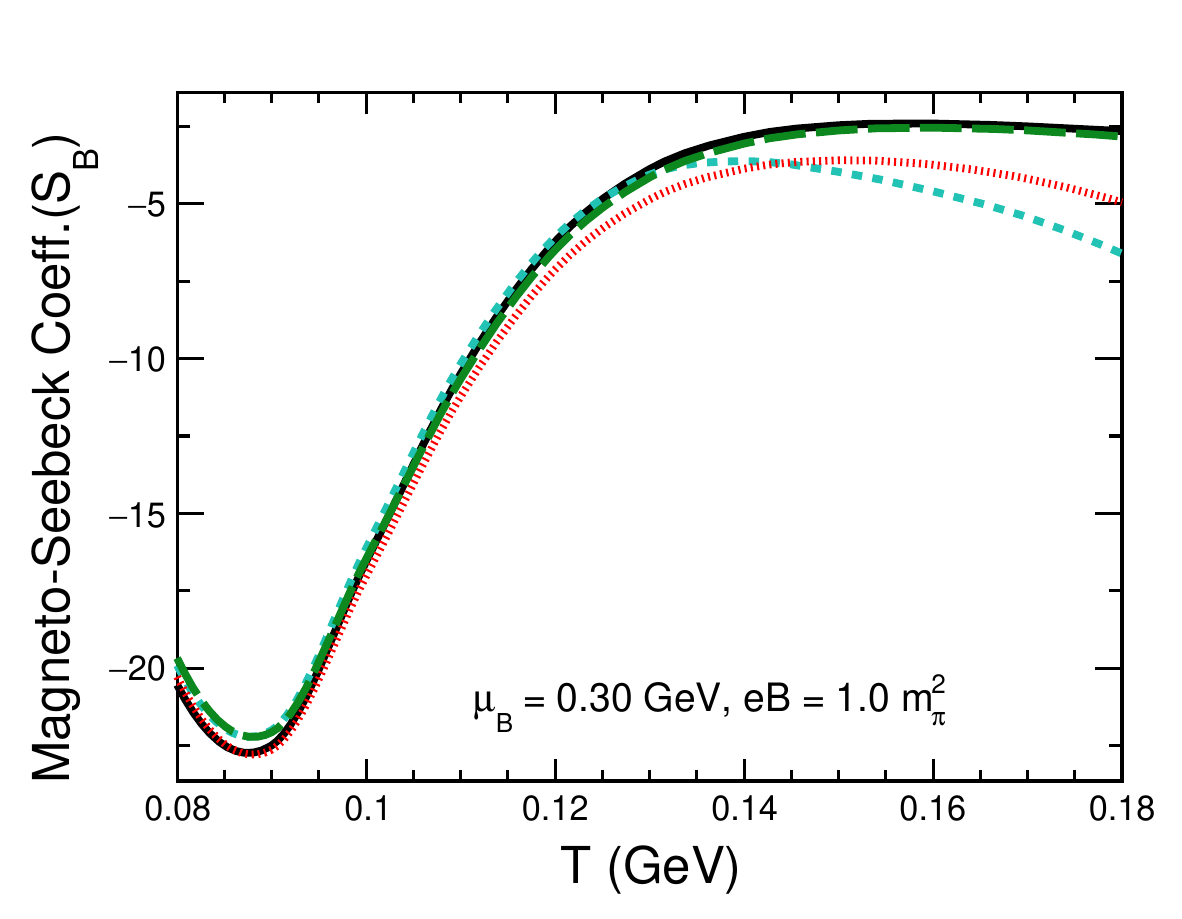}
    \includegraphics[scale=0.27]{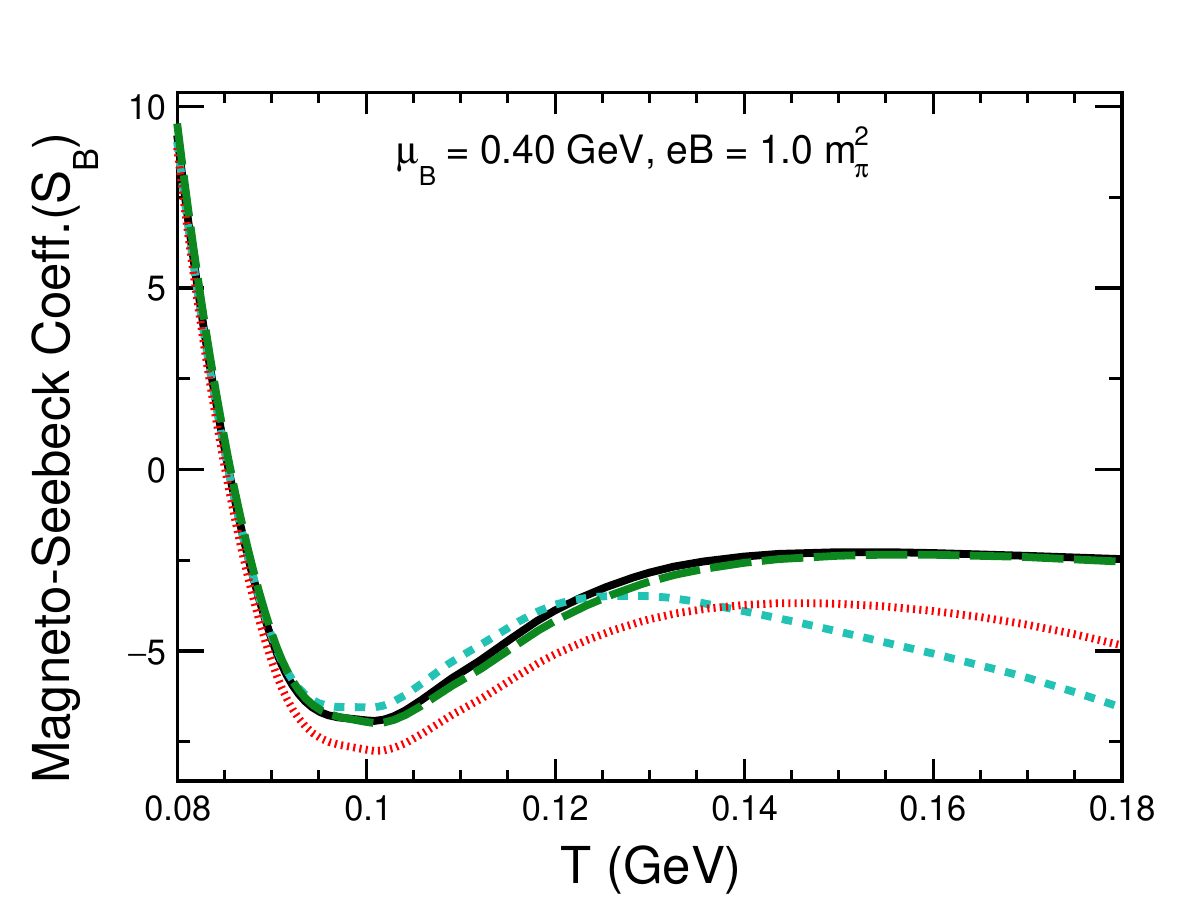}
    
	\caption{Magneto-Seebeck coefficient ($S_{B}$) as a function of temperature at magnetic field $eB$ = 0.1 $m_{\pi}^2$ (upper panel) and $eB$ = 1.0 $m_{\pi}^2$ (lower panel) for baryon chemical potential at $\mu_{B}$ 0.10 GeV (left), 0.30 GeV (middle), and 0.40 GeV (right).}
	\label{Fig-magseebeck1}
\end{figure*}

In Fig.~\ref{Fig-seecom}, we compare our results of the Seebeck coefficient with those obtained in different phenomenological works at $\mu_B$ = 0.10 GeV. The $S$ obtained in Ref.~\cite{Das:2020beh} using the IHRG model (Magenta dotted line) is found to be less in magnitude as compared to our results. This can be attributed to the use of different hadronic radii for the hard-sphere scattering cross section to estimate the relaxation time. In Ref.~\cite{Das:2020beh}, a uniform radius, $r_h$ = 0.3 fm, is taken for all the hadrons, whereas we choose $r_M$ = 0.2 fm for mesons and $r_B$ = 0.62 fm for (anti)baryons, which give a better description for mesons and baryons. The result from the Nambu–Jona Lasinio (NJL) model~\cite{Abhisek2022} for the confined phase is shown in a dotted dashed blue line.

Fig.~\ref{Fig-Thomson1} represents the Thomson coefficient, $Th$ as a function of temperature for the same baryon chemical potentials, i.e., at $\mu_{B}$ = 0.10, 0.30, and 0.40 GeV. In condensed matter systems, the Thomson coefficient for a system refers to the absorption or release of heat ($Q$) in the system when a current $\vec{j}$ flows through it corresponding to the temperature gradients $\vec{\nabla} T$, as $Q = Th ~\vec{j}\cdot\vec{\nabla} T$ \cite{ioffe1957physics}. The positive (negative) values of the Thomson coefficient lead to the absorption (release) of the heat in the system. Therefore, the presence of a nonzero Thomson coefficient may affect the cooling rate of the system. From fig.~\ref{Fig-Thomson1} we observe that similar to the Seebeck coefficient, Th also gives negative values for all three cases, decreasing universally as a function of temperature. As given in Eq.~(\ref{Thomson}), the Thomson coefficient is also defined as the product of temperature to the rate of change of the Seebeck coefficient with respect to temperature. So, the higher the steepness of the  Seebeck coefficient in Fig.~\ref{Fig-seebeck1}, the higher the value of $Th$ in magnitude as a function of $T$. Therefore, in a similar behavior to the Seebeck coefficient, the value of $Th$ is found to be more negative (i.e., more amount of heat is released) towards high temperature for EVHRG and VDWHRG. Whereas all these models overlap at lower values of temperature nearly up to $\approx 0.14$ GeV for the case of $\mu_B$ = 0.10 GeV. As the value of $\mu_B$ increases, the models start deviating from each other even at low temperatures, as in the case of the Seebeck coefficient, which is explained above.

To study the efficiency of QGP medium as a thermoelectric medium, we draw the thermoelectric figure of merit, $ZT$ as a function of temperature in Fig.~\ref{Fig-ZT}. As given in Eq.~(\ref{eqnZT}), $ZT$ can be defined as the ratio of $S^2$ to the Lorenz number ($L$), which is proportional to the ratio of thermal to electrical conductivity and defined as $L = \kappa_0/\sigma_{el} T$~\cite{Pradhan:2022gbm, Singh:2023pwf}. Since it is proportional to $S^2$, we observed that $ZT$ increases both as a function of $T$ and $\mu_{B}$. The rise in $\mu_{B}$ gives rise to $ZT$ because of the presence of more baryons as compared to antibaryons. Hence, the net thermoelectric current increases, which ultimately improves the thermoelectric efficiency of the medium. The sudden increase of ZT for the VDWHRG case can be attributed to the rise of pressure as well as number density, which subsequently affects other transport properties, due to the interplay between both attractive and repulsive interactions between the hadrons. However, the decreasing trend in the VDWHRG case may be due to the effect of liquid-gas phase transition, which comes at a lower temperature with increasing baryochemical potential.

 \begin{figure*}
	\centering
	\includegraphics[scale=0.27]{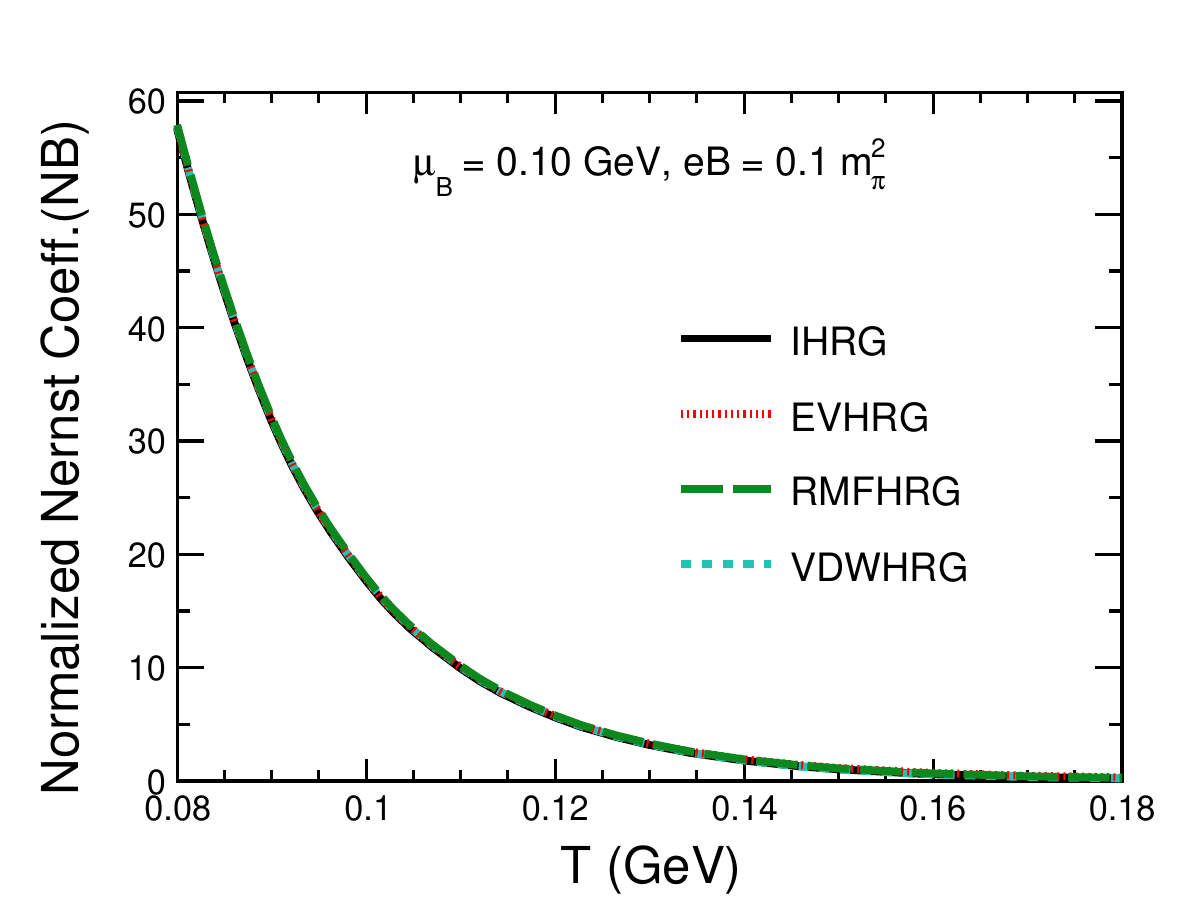}
    \includegraphics[scale=0.27]{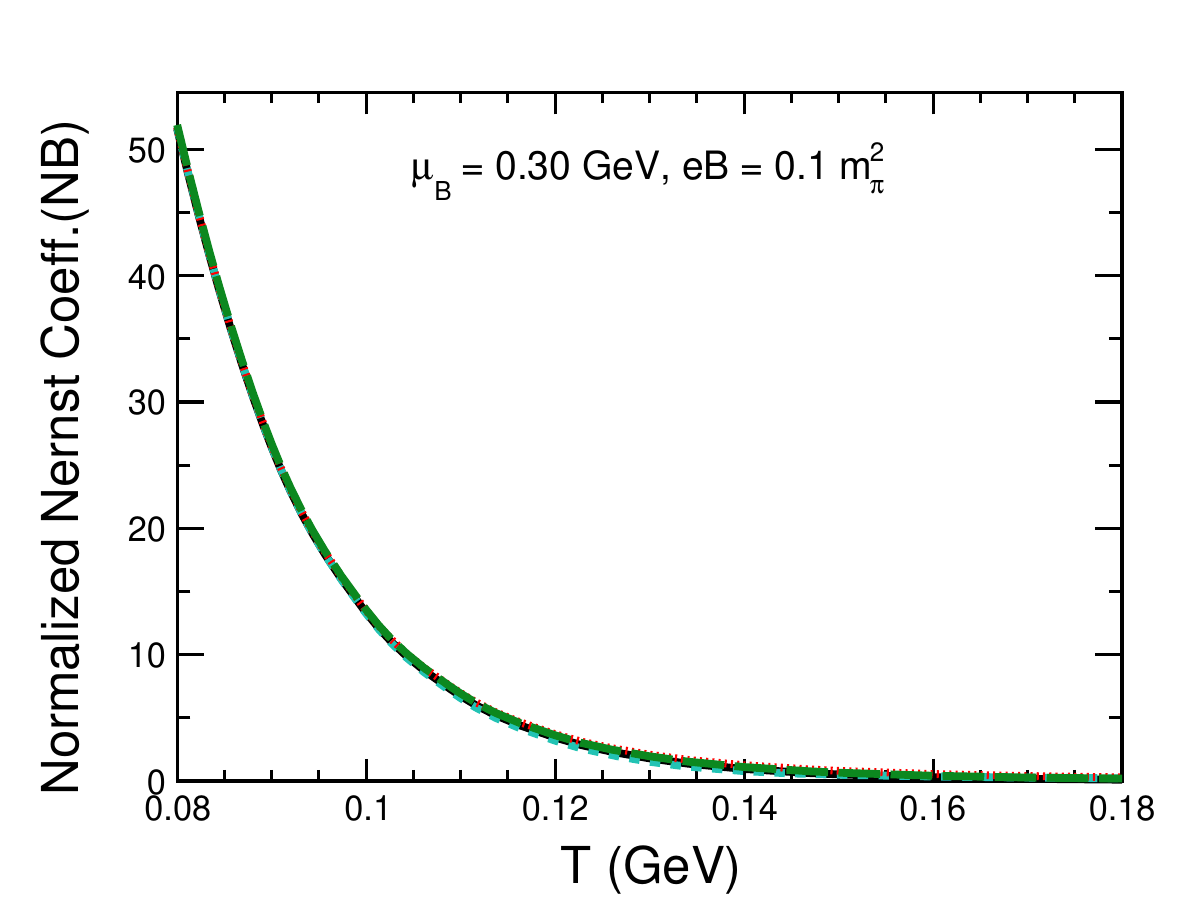}
    \includegraphics[scale=0.27]{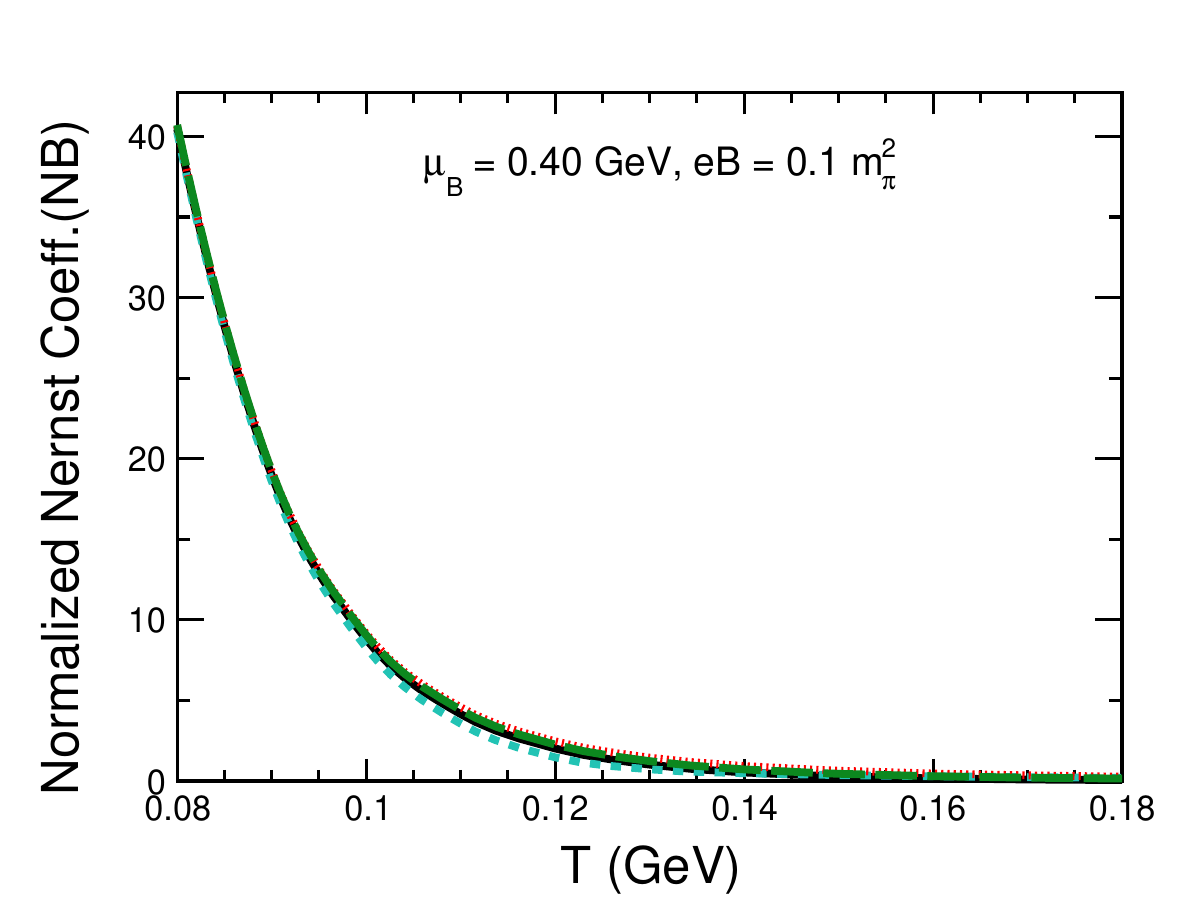}

	\includegraphics[scale=0.27]{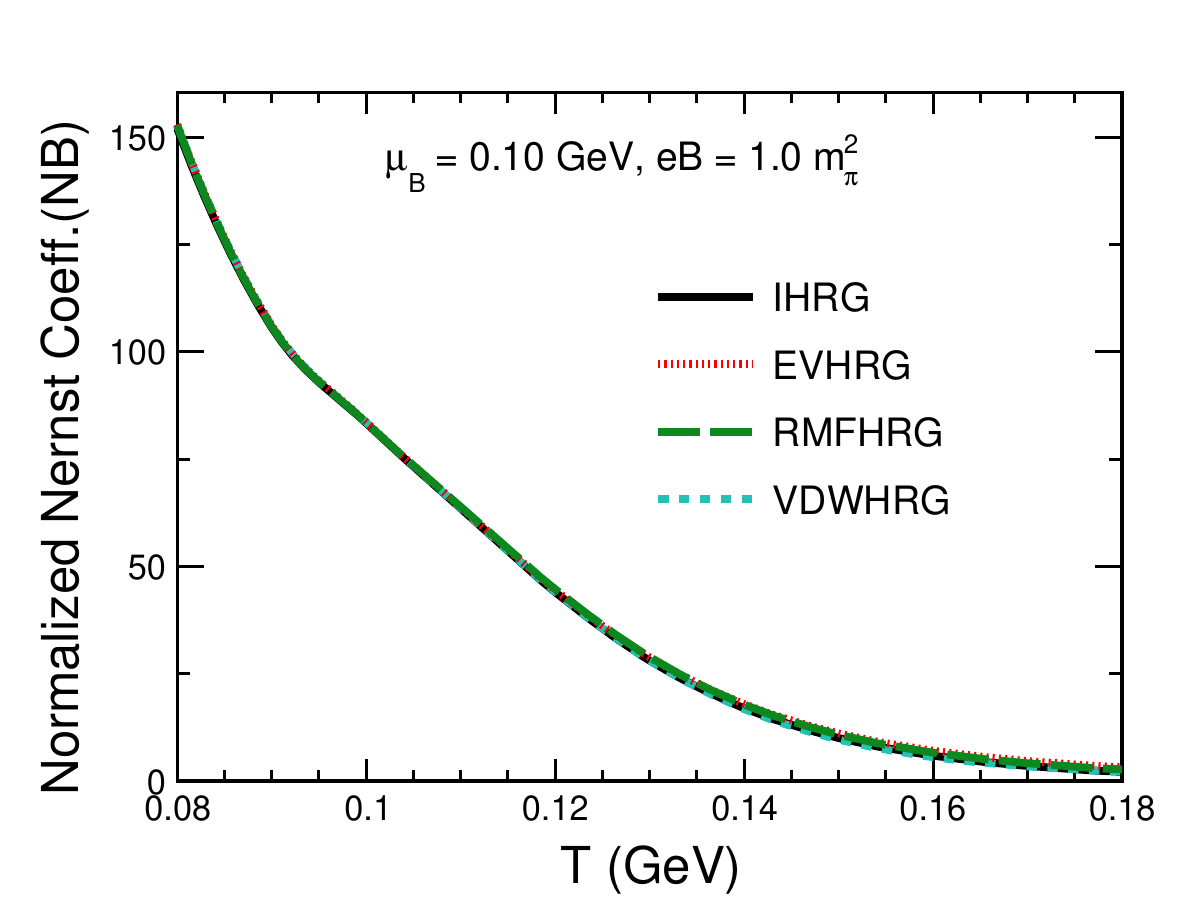}
    \includegraphics[scale=0.27]{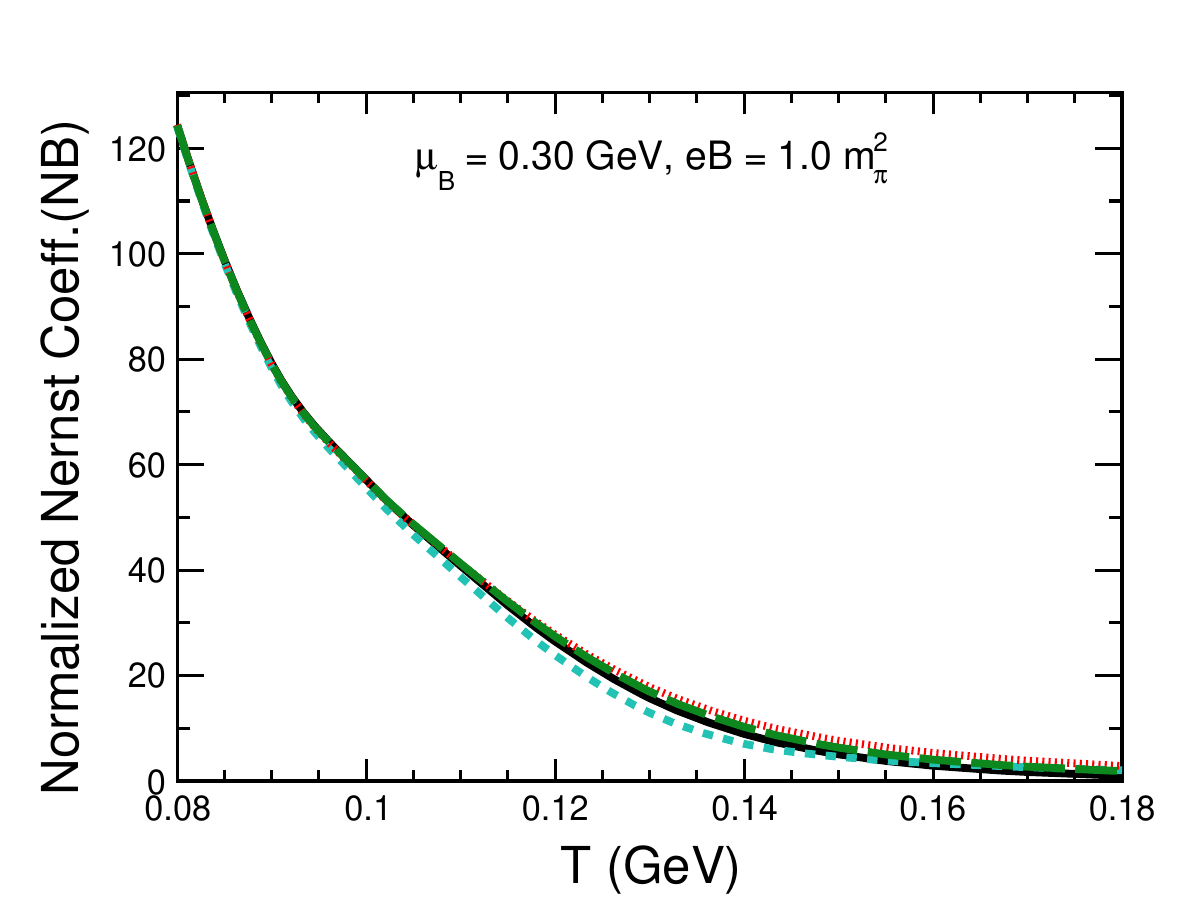}
    \includegraphics[scale=0.27]{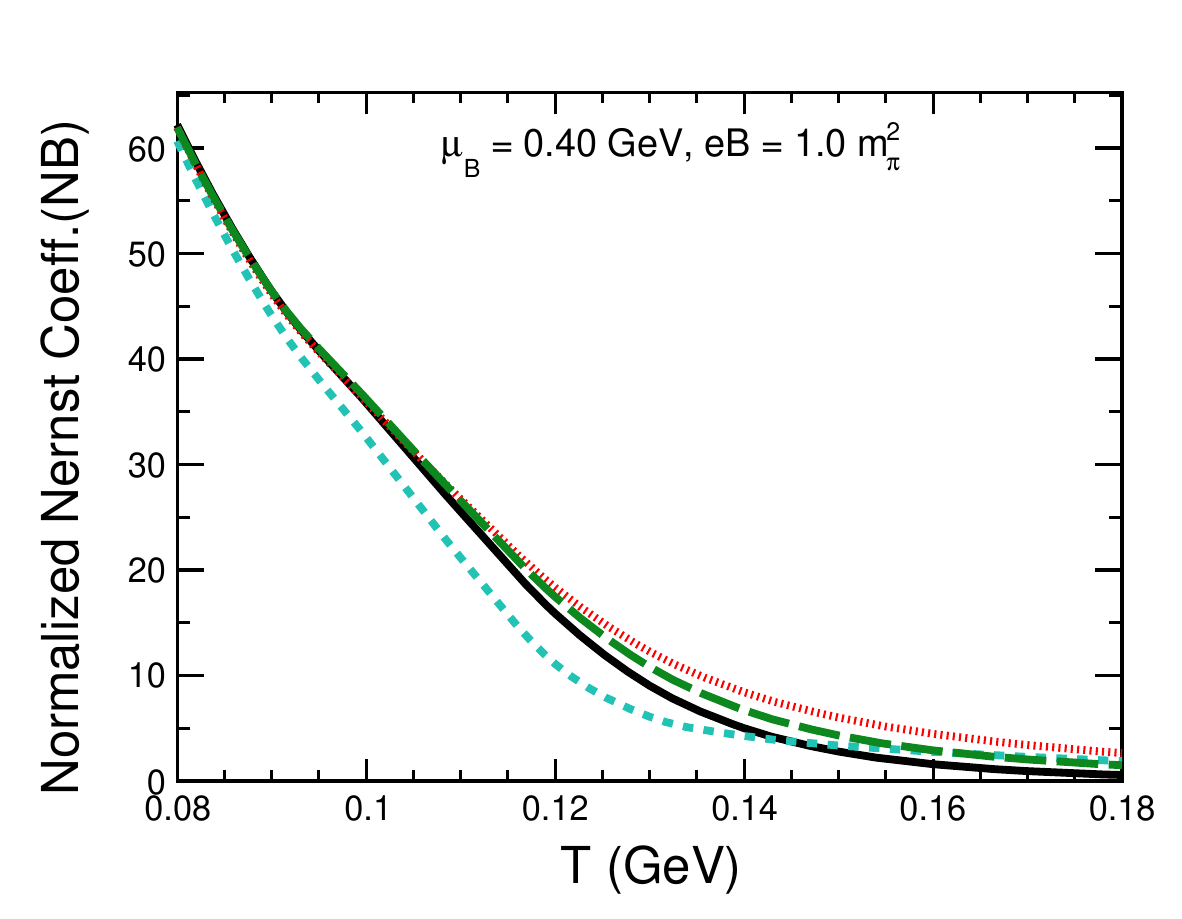}
	\caption{Normalized Nernst coefficient ($NB$) as a function of temperature at magnetic field $eB$ = 0.1 $m_{\pi}^2$ (upper panel) and $eB$ = 1.0 $m_{\pi}^2$ (lower panel) for baryon chemical potential at $\mu_{B}$ 0.10 GeV (left), 0.30 GeV (middle), and 0.40 GeV (right).}
	\label{Fig-nernst1}
\end{figure*}

\begin{figure*}
	\centering
	\includegraphics[scale=0.27]{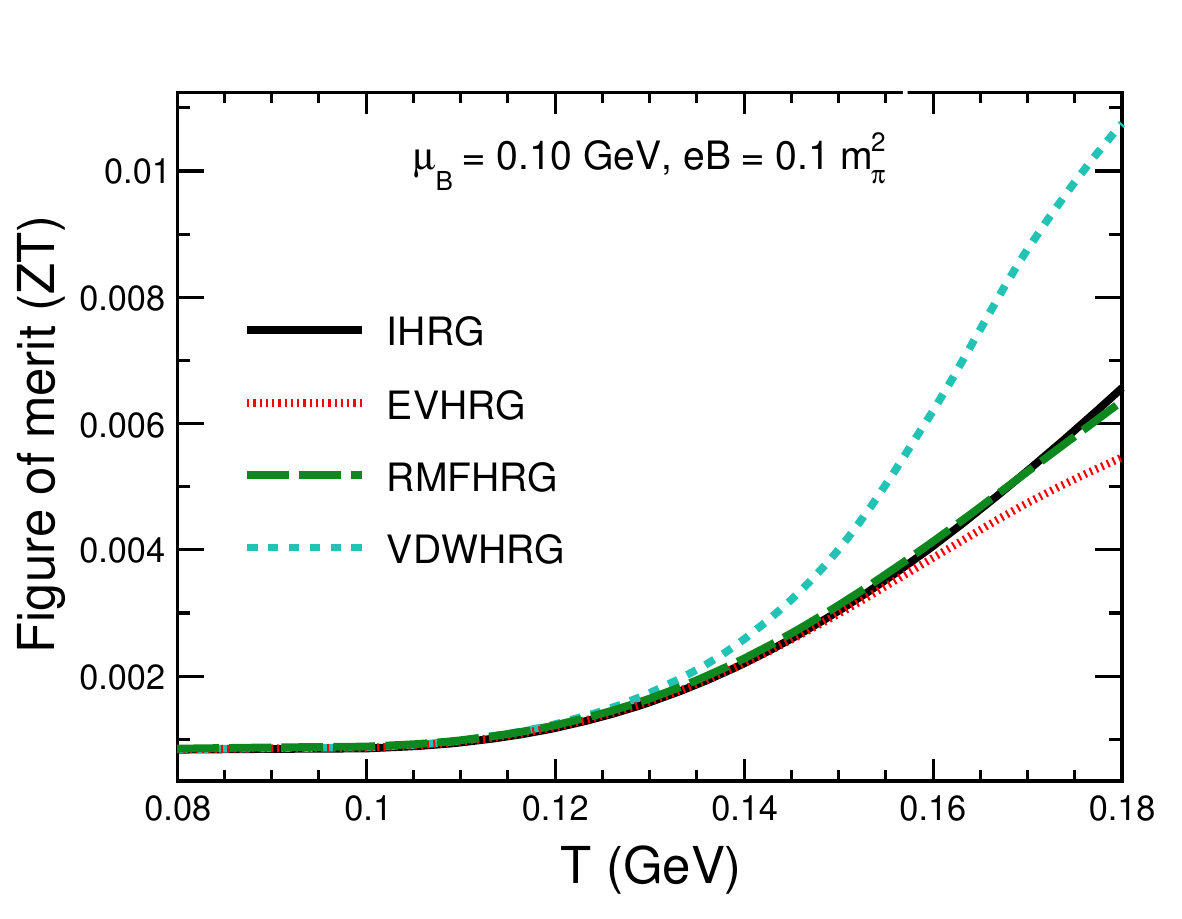}
    \includegraphics[scale=0.27]{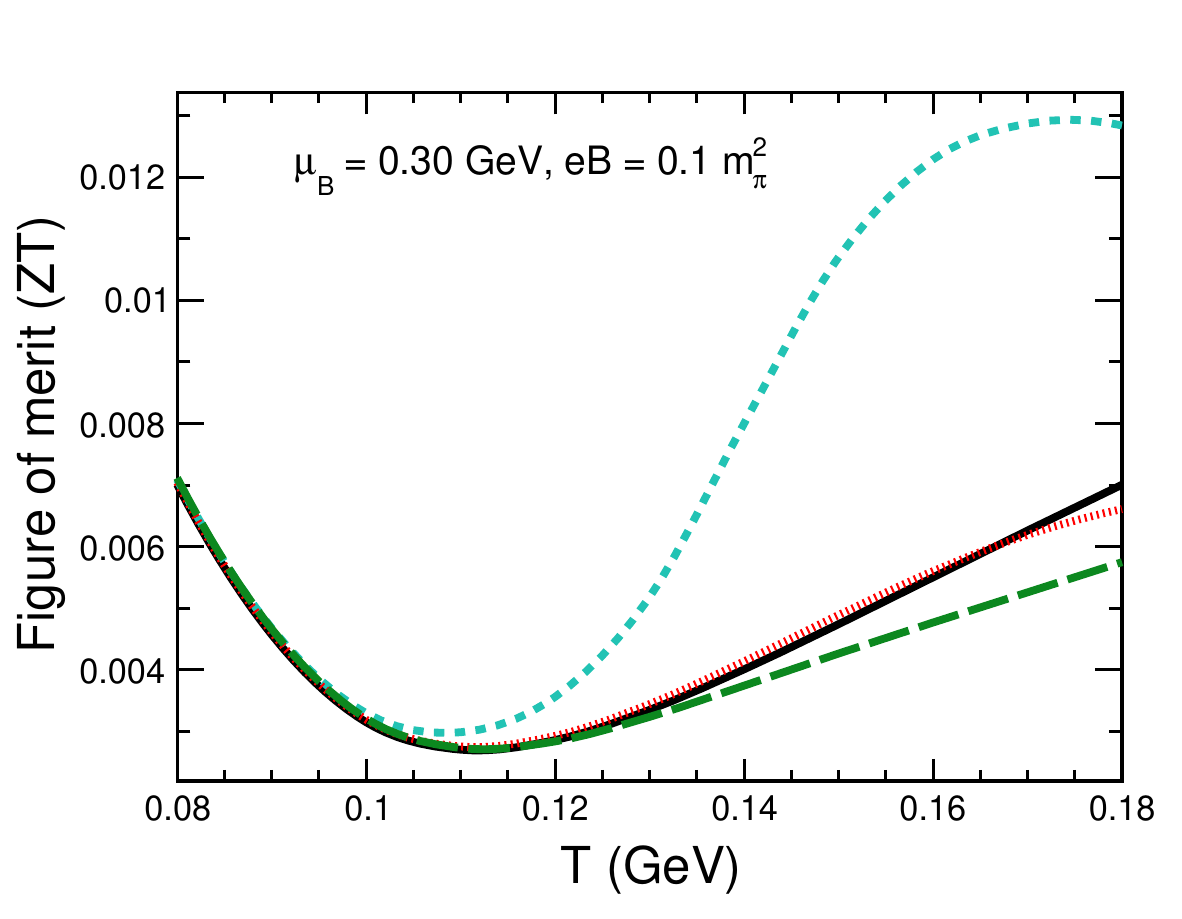}
    \includegraphics[scale=0.27]{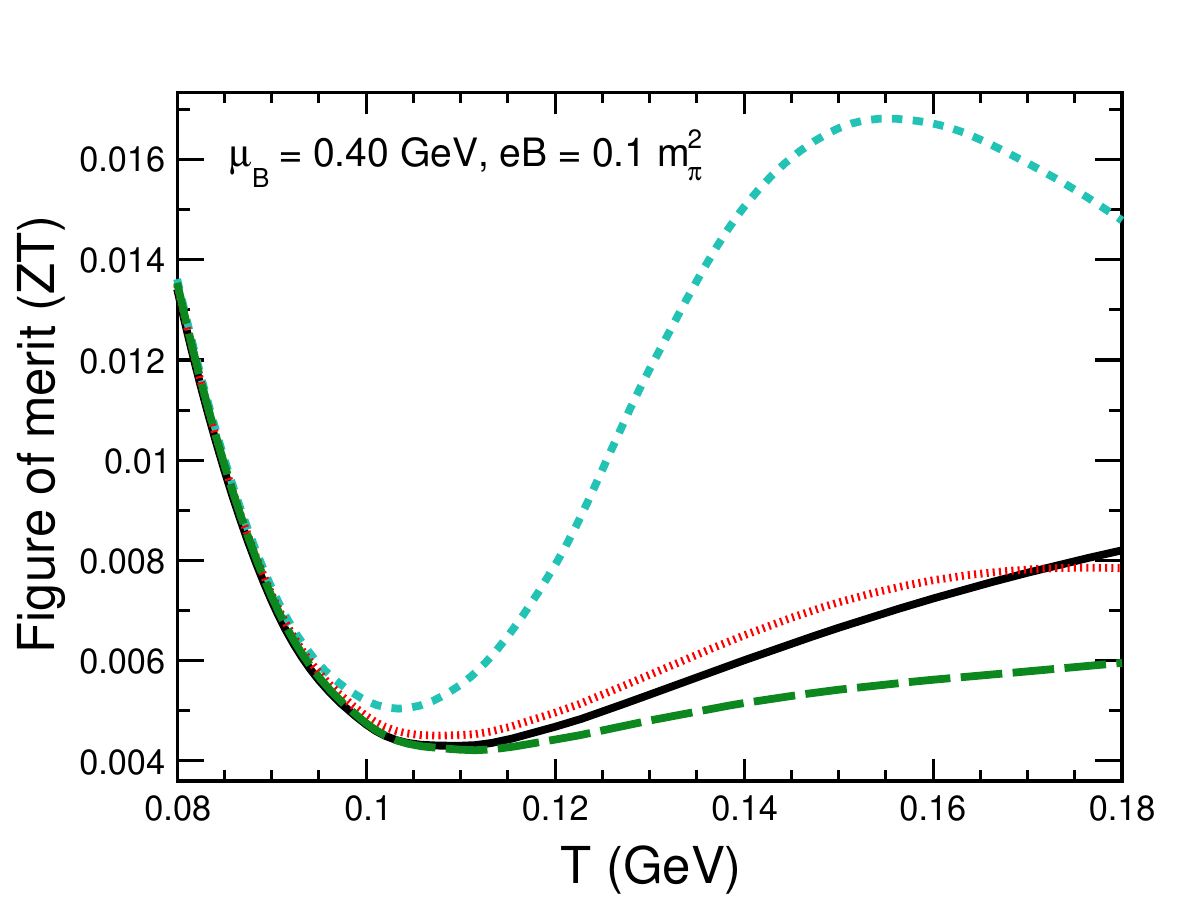}

	\includegraphics[scale=0.27]{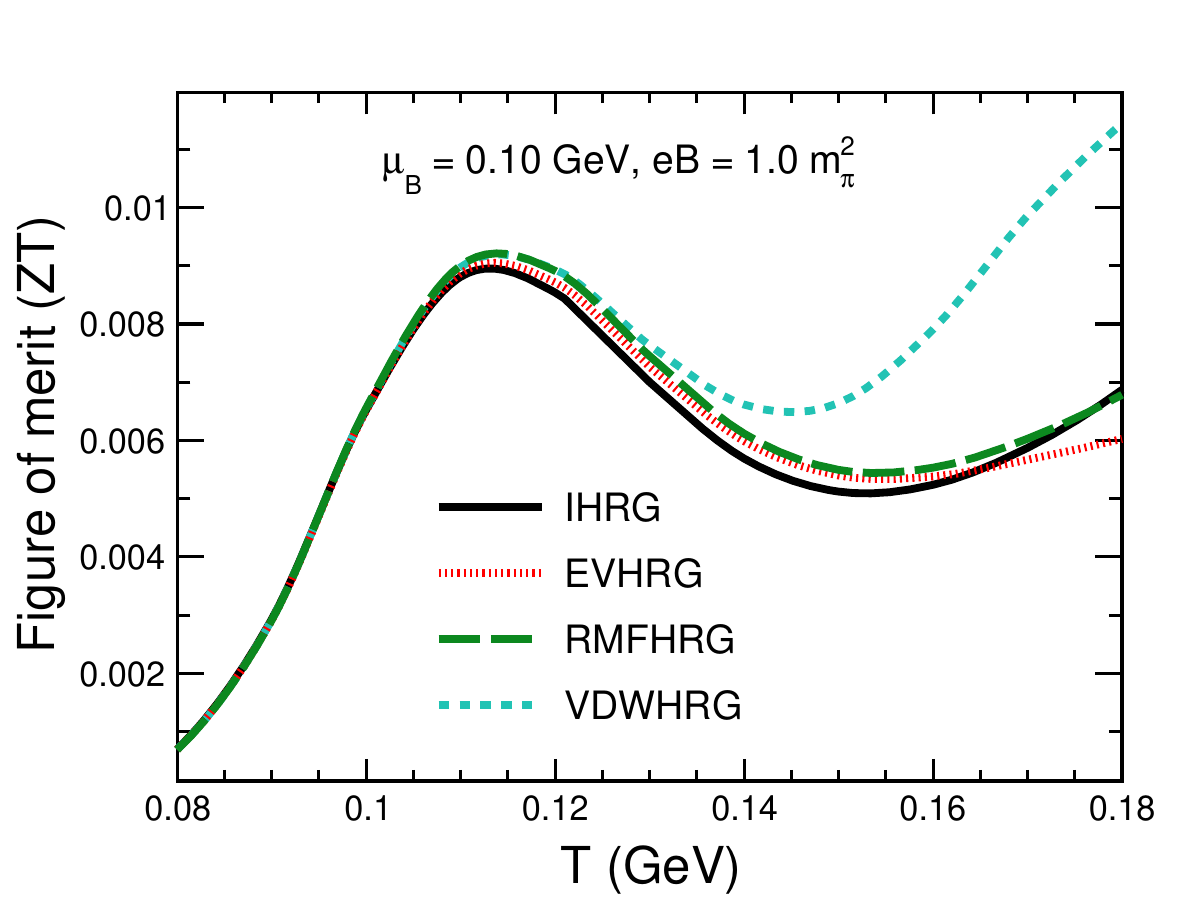}
    \includegraphics[scale=0.27]{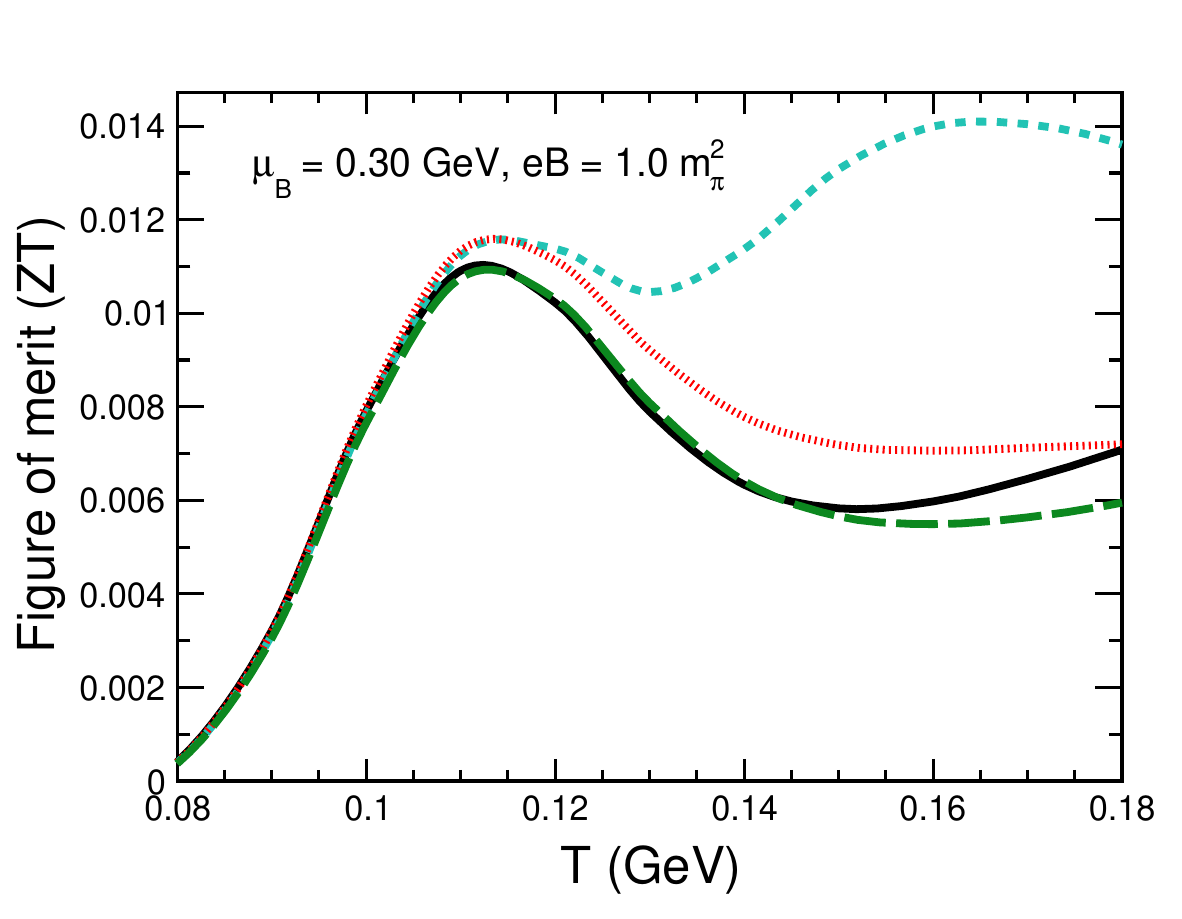}
    \includegraphics[scale=0.27]{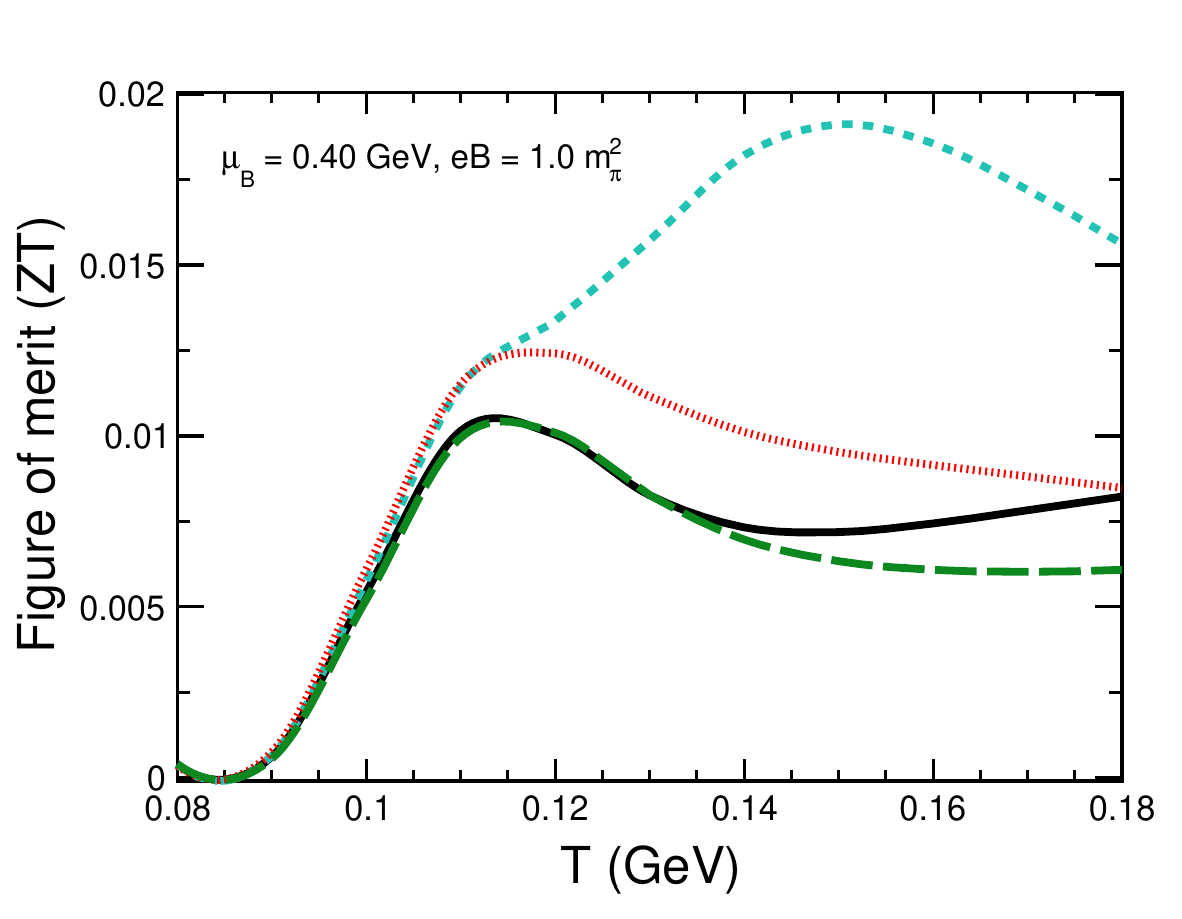}
	\caption{Thermoelectric figure of merit ($ZT$) as a function of temperature at magnetic field $eB$ = 0.1 $m_{\pi}^2$ (upper panel) and $eB$ = 1.0 $m_{\pi}^2$ (lower panel) for baryon chemical potential at $\mu_{B}$ 0.10 GeV (left), 0.30 GeV (middle), and 0.40 GeV (right).}
	\label{Fig-ZT1}
\end{figure*}

\subsection{In presence of magnetic field}
Here, we discuss the magneto-Seebeck coefficient, $S_{B}$ in the presence of a nonzero magnetic field as a function of $T$ for different values of $\mu_{B}$. The upper panels of Fig.~\ref{Fig-magseebeck1} represent $S_{B}$ as a function of $T$ for three different values of $\mu_{B}$ at  0.10, 0.30, and 0.40 GeV at magnetic field 0.1 $m_{\pi}^2$, whereas the lower panels represent those for the magnetic field 1.0 $m_{\pi}^2$. For the case of $eB$ = 0.1 $m_\pi^2$, we observe that $S_{B}$ increases with the temperature almost linearly up to 0.1 GeV and then starts to decrease at higher temperatures. Similar to what was observed in the case of a vanishing magnetic field, all the models agree up to a certain temperature and then start deviating from each other at higher temperatures. Furthermore, we also observe that the values of $S_{B}$ increase with the increasing values of $\mu_{B}$. In contrast to the zero magnetic field case, as shown in Eq.~(\ref{equnew49}), $S_{B}$ has a dependence on the Hall-like components of electrical conductivity and other integral terms which comes into the picture because of the presence of the external magnetic field. The Hall-like component of electrical conductivity, $\sigma_{H}/T$, increases with $\mu_{B}$ because of an increase in net baryonic contributions. It is also to be noted here that there is no net contribution of mesons in the integral $\mathcal{I}_{31}/T^2$ and $\mathcal{I}_{42}/T^2$, but it enters through enthalpy per baryon ($h$). The effect of the magnetic field is significant on $\mathcal{I}_{31}/T^2$ at lower temperatures, but for higher temperatures, it is not effective due to the lower value of relaxation time. We also observe that $\mathcal{I}_{42}/T^2$ first increases with temperature and then starts to decrease. Higher is the magnetic field, and lower is the $\mathcal{I}_{42}/T^2$. The order of magnitude of $\sigma_{el}/T$ is larger than $\sigma_{H}/T$, whereas the order of magnitude of $\mathcal{I}_{31}/T^2$ is close to $\mathcal{I}_{42}/T^2$. Hence, the behavior of the magneto-Seebeck coefficient can be approximated as the ratio of $(\mathcal{I}_{31}/T^2)/(\sigma_{el}/T)$ which can be seen in Fig.~\ref{Fig-magseebeck1}. The dependency of $\sigma_{el}/T$ on $T$ and $\mu_B$ in the presence of a magnetic field results in first increase of $S_B$ at low $T$ and then decrease at high $T$. This is because, in the presence of a magnetic field, the higher value of $\Omega_{c_{i}}\tau_{R_{i}}$ at the lower temperature range affects the mesonic contribution to $\sigma_{el}/T$ significantly. This decrease in mesonic contribution gets compensated for by an increase in baryon contribution with increasing $\mu_{B}$. On the other hand, this decrease in the mesonic contribution is not compensated by those increasing baryonic contributions for higher temperature ranges. Hence, $\sigma_{el}/T$ shows decreasing trend with increasing $\mu_{B}$ for higher temperatures. For a higher value of the magnetic field as shown in the lower panels in Fig.~\ref{Fig-magseebeck1}, we observe a minimum for $S_B$ while going to the high $\mu_B$ region. This minimum shifts towards high $T$ when $\mu_B$ increases further.

In the presence of an external magnetic field, one can also expect the Hall-type thermoelectric coefficient. If there is a temperature gradient in a conducting medium perpendicular to the direction of a magnetic field, then the Nernst coefficient gives the measure of electric current perpendicular to both the magnetic field and temperature gradient. In the absence of any magnetic field, this quantity vanishes. Here, we discuss the normalized Nernst coefficient, $NB$ as a function of $T$ and $\mu_{B}$ in the presence of a finite magnetic field. Fig.~\ref{Fig-nernst1} represents $NB$ as a function of $T$ for three different values of $\mu_{B}$ at  0.10, 0.30, and 0.40 GeV. The upper and lower panel presents the results for the case of the magnetic field with values 0.1 and 1.0 $m_{\pi}^2$, respectively. In both figures, we observe that $NB$ decreases with increasing temperature as well as with baryon chemical potential. For $eB$ = 0.1~$m_{\pi}^2$, not much difference between the results from different models is observed. Even at higher $\mu_B$, only a slight deviation in the models can be seen. As one goes to a higher value of the magnetic field, for example, at $eB$ = 1.0~$m_\pi^2$, $NB$ is found to be higher. Here also, $NB$ nearly overlaps for all the models at lower temperatures, but the deviation is observed at higher temperatures and high $\mu_B$. Due to the higher order of magnitude of $\sigma_{el}/T$ as compare to $\sigma_{H}/T$, $NB$ can be approximated by ratio $(\mathcal{I}_{42}/T^2)/(\sigma_{el}/T)$. The sharp decrease of $\mathcal{I}_{42}/T^2$ as compare to $\sigma_{el}/T$ give rise to decreasing trend of $NB$ with $\mu_{B}$. 

Finally, we discuss the variation of the figure of merit with temperature and baryon chemical potential in the presence of an external magnetic field. Fig.~\ref{Fig-ZT1} shows $ZT$ as a function of $T$ for three different values of $\mu_{B}$ at  0.10, 0.30, and 0.40 GeV at two different magnetic field orders, $eB$ = 0.1 $m_{\pi}^2$ in the upper panels, whereas the lower panels represent those for $eB$ = 1.0 $m_{\pi}^2$. In the presence of a magnetic field, the behavior of $ZT$ is different at low temperatures as compared to those in the case of vanishing magnetic fields. At higher temperatures, the results are nearly the same as in the case of the vanishing magnetic field, as shown in Fig.~\ref{Fig-ZT}. As the value of $\mu_B$ increases, a minimum appears at low temperatures. For the case of $eB$ = 1.0 $m_{\pi}^2$, $ZT$ gets fully affected throughout the temperature range considered. This is because $ZT$ is proportional to the square of the magneto-Seebeck coefficient, and hence, the effect of the magnetic field on $S_B$ is reflected in $ZT$. The presence of magnetic field results in the highly non-monotonic behavior of $ZT$ as a function of both $T$ and $\mu_B$. We observe that the values of the thermoelectric figure of merit overlap for all the models at low temperatures, but a large deviation is observed at high temperatures. Along with the Ohmic-like component of thermal conductivity ($\kappa_0$), the Hall-like component of thermal conductivity ($\kappa_H$)~\cite{Singh:2023pwf} also contributes to $ZT$ in the presence of an external magnetic field. However, the effect of $\kappa_H$ becomes more significant in the presence of strong magnetic field.

\section{Summary}
\label{summary}
The presence of temperature gradients from the central to the peripheral region of heavy-ion collision can lead to thermoelectric currents in the medium in the case of nonzero baryonic chemical potential. We have calculated several thermoelectric coefficients within the ideal and other interacting HRG models. For the first time for a hot and dense hadronic medium, we calculated the Thomson coefficient, which was introduced due to the temperature dependence of the Seebeck coefficient. For the completeness of our study, we have also calculated the thermoelectric figure of merit $ZT$ of the hadronic medium. We relate the gradients of baryon chemical potential to the gradients of temperature using the Gibbs-Duhem relation. During this modification, we observed the introduction of enthalpy in the Seebeck coefficient formula. Although the Seebeck coefficient $S$ for mesons vanishes due to equal and opposite charges, mesons still have a significant effect on the $S$ because of their contributions to enthalpy. 
We have also studied the effects of finite magnetic fields on the thermoelectric properties of the HRG medium. As the hadronic medium is a conducting medium, the presence of a magnetic field leads to Hall-like components due to the Lorentz force. In the absence of a magnetic field, the magneto-Seebeck coefficient $S_B$ merges into the Seebeck coefficient. We observed negative values $S_{B}$ for all cases, except for the high magnetic field ($eB = 1.0~ m_{\pi}^{2}$) and baryon chemical potential ($\mu_{B}$ = 0.40 GeV) at lower temperatures where $S_{B}$ is positive. On the other hand, the induced current in the direction perpendicular to the electric field is measured with the help of the normalized Nernst coefficient $NB$. The higher the magnetic field value, the higher the value of $NB$. For all the cases, we observe that at higher temperatures, $NB$ approaches zero. The presence of a magnetic field also affects the thermoelectric figure of merit $ZT$. The higher the magnetic field, the higher the nonmonotonic behavior of $ZT$. The thermoelectric efficiency of the medium is the same for all the HRG models at the lower temperature with finite values of the magnetic field at a fixed $\mu_{B}$. However, it is highly deviated for each of the models at higher temperatures.  

The current study focuses on estimating the thermoelectric coefficients of a hot and dense hadronic medium formed in heavy-ion collisions using different HRG models to study the impact of hadronic interactions. However, it will also be interesting to study the magneto-Thomson effect in hadron gas and the QGP medium. 

\section*{ACKNOWLEDGMENTS}
K.S and K.K.P. acknowledge the financial aid from UGC, Government of India. The authors gratefully acknowledge the DAE-DST, Govternment of India funding under the mega-science project – “Indian participation in the ALICE experiment at CERN” bearing Project No. SR/MF/PS-02/2021-
IITI (E-37123). 


 \end{document}